\documentclass[structabstract]{aa}  
\usepackage{graphicx}
\usepackage{txfonts}
\usepackage{color}
\usepackage{deluxetable}
\definecolor{hxcol}{rgb}{0.8,0.2,0.2}
\definecolor{normal}{rgb}{0.0,0.0,0.0}

\newcommand{\normal}{\color{normal}}

\newcommand{\keyw}[1]{\mbox{{\tt #1}}}

\newcommand{\keyi}[2]{\mbox{{\tt #1\hspace{1pt}}{$#2$}\/}}
\newcommand{\CRPIX}[1]{\keyi{CRPIX}{#1}}
\newcommand{\CDELT}[1]{\keyi{CDELT}{#1}}

\newcommand{\CRVAL}[1]{\keyi{CRVAL}{#1}}
\newcommand{\CNAME}[1]{\keyi{CNAME}{#1}}
\newcommand{\CTYPE}[1]{\keyi{CTYPE}{#1}}
\newcommand{\PTYPE}[1]{\keyi{PTYPE}{#1}}
\newcommand{\CUNIT}[1]{\keyi{CUNIT}{#1}}
\newcommand{\DATEREF}[1]{\keyi{DATEREF}{#1}}
\newcommand{\JDREF}[1]{\keyi{JDREF}{#1}}

\newcommand{\MJDREF}[1]{\keyi{MJDREF}{#1}}

\newcommand{\TCDE}[1]{\keyi{TCDE}{#1}}
\newcommand{\TCDLT}[1]{\keyi{TCDLT}{#1}}
\newcommand{\TCRV}[1]{\keyi{TCRV}{#1}}
\newcommand{\TCRVL}[1]{\keyi{TCRVL}{#1}}
\newcommand{\TCRP}[1]{\keyi{TCRP}{#1}}
\newcommand{\TCRPX}[1]{\keyi{TCRPX}{#1}}
\newcommand{\TCNA}[1]{\keyi{TCNA}{#1}}
\newcommand{\TCNAM}[1]{\keyi{TCNAM}{#1}}
\newcommand{\TCTYP}[1]{\keyi{TCTYP}{#1}}
\newcommand{\TCTY}[1]{\keyi{TCTY}{#1}}
\newcommand{\TCUN}[1]{\keyi{TCUN}{#1}}
\newcommand{\TCUNI}[1]{\keyi{TCUNI}{#1}}
\newcommand{\TCSY}[1]{\keyi{TCSY}{#1}}
\newcommand{\TCSYE}[1]{\keyi{TCSYE}{#1}}
\newcommand{\CSYER}[1]{\keyi{CSYER}{#1}}
\newcommand{\TCRD}[1]{\keyi{TCRD}{#1}}
\newcommand{\TCRDE}[1]{\keyi{TCRDE}{#1}}
\newcommand{\CRDER}[1]{\keyi{CRDER}{#1}}
\newcommand{\TFORM}[1]{\keyi{TFORM}{#1}}

\newcommand{\TIMEDEL}[1]{\keyi{TIMEDEL}{#1}}

\newcommand{\TIMESYS}[1]{\keyi{TIMESYS}{#1}}
\newcommand{\TIMEUNIT}[1]{\keyi{TIMEUNIT}{#1}}

\newcommand{\TREFPOS}[1]{\keyi{TREFPOS}{#1}}
\newcommand{\TRPOS}[1]{\keyi{TRPOS}{#1}}
\newcommand{\TREFDIR}[1]{\keyi{TREFDIR}{#1}}
\newcommand{\TRDIR}[1]{\keyi{TRDIR}{#1}}

\newcommand{\TTYPE}[1]{\keyi{TTYPE}{#1}}

\newcommand{\CZPHS}[1]{\keyi{CZPHS}{#1}}
\newcommand{\TCZPH}[1]{\keyi{TCZPH}{#1}}
\newcommand{\TCZP}[1]{\keyi{TCZP}{#1}}
\newcommand{\CPERI}[1]{\keyi{CPERI}{#1}}
\newcommand{\TCPER}[1]{\keyi{TCPER}{#1}}
\newcommand{\TCPR}[1]{\keyi{TCPR}{#1}}

\newcommand{\keyii}[3]{\mbox{{\tt #1\hspace{1pt}{$#2$}\_{$#3$}}\/}}
\newcommand{\PC}[2]{\keyii{PC}{#1}{#2}}
\newcommand{\TP}[2]{\keyii{TP}{#1}{#2}}
\newcommand{\CD}[2]{\keyii{CD}{#1}{#2}}
\newcommand{\TC}[2]{\keyii{TC}{#1}{#2}}

\newcommand{\keyij}[3]{\mbox{{\tt {$#2$}\hspace{1pt}{#1}\hspace{1pt}{$#3$}}\/}}
\newcommand{\iPC}[2]{\keyij{PC}{#1}{#2}}
\newcommand{\iCD}[2]{\keyij{CD}{#1}{#2}}
\newcommand{\iCTYP}[2]{\keyij{CTYP}{#1}{#2}}
\newcommand{\iCTY}[2]{\keyij{CTY}{#1}{#2}}
\newcommand{\iCUNI}[2]{\keyij{CUNI}{#1}{#2}}
\newcommand{\iCUN}[2]{\keyij{CUN}{#1}{#2}}
\newcommand{\iCSYE}[2]{\keyij{CSYE}{#1}{#2}}
\newcommand{\iCSY}[2]{\keyij{CSY}{#1}{#2}}
\newcommand{\iCRDE}[2]{\keyij{CRDE}{#1}{#2}}
\newcommand{\iCRD}[2]{\keyij{CRD}{#1}{#2}}
\newcommand{\iCNAM}[2]{\keyij{CNAM}{#1}{#2}}
\newcommand{\iCNA}[2]{\keyij{CNA}{#1}{#2}}
\newcommand{\iCRPX}[2]{\keyij{CRPX}{#1}{#2}}
\newcommand{\iCRP}[2]{\keyij{CRP}{#1}{#2}}
\newcommand{\iCRVL}[2]{\keyij{CRVL}{#1}{#2}}
\newcommand{\iCRV}[2]{\keyij{CRV}{#1}{#2}}
\newcommand{\iCDLT}[2]{\keyij{CDLT}{#1}{#2}}
\newcommand{\iCDE}[2]{\keyij{CDE}{#1}{#2}}
\newcommand{\iCPER}[2]{\keyij{CPER}{#1}{#2}}
\newcommand{\iCPR}[2]{\keyij{CPR}{#1}{#2}}
\newcommand{\iCZPH}[2]{\keyij{CZPH}{#1}{#2}}
\newcommand{\iCZP}[2]{\keyij{CZP}{#1}{#2}}

\newcommand{\keyv}[1]{\mbox{{\tt #1}}}

\newcommand{\newkey}[2]{\begin{quote}\centering #1 (#2-valued)\end{quote}}
\newcommand{\newkeyp}[3]{\begin{quote} \textbf{#1} (#2-valued)\\
    \small #3\end{quote}}

\begin{document}
   \title{Representations of Time Coordinates in FITS}
   \subtitle{Time and Relative Dimension in Space}

   \author{Arnold H.\ Rots
          \inst{1}
	  \and
          Peter S.\ Bunclark
	  \inst{2}
	  \fnmsep
	  \thanks{Deceased}
          \and
          Mark R.\ Calabretta
	  \inst{3}
	  \and
          Steven L.\ Allen
	  \inst{4}
          \and
          Richard N.\ Manchester
	  \inst{3}
          \and
          William T.\ Thompson
	  \inst{5}
          }

   \institute{Harvard-Smithsonian Center for Astrophysics/Smithsonian
             Astrophysical Observatory,
             60 Garden Street MS 67, Cambridge, MA 02138, USA;
             arots@cfa.harvard.edu\\
         \and
             Institute of Astronomy, Madingley Road, Cambridge
             CB3 0HA, UK\\
	 \and
	     CSIRO Astronomy and Space Science,
	     PO Box 76, Epping, NSW 1710, Australia\\
         \and
             UCO/Lick Observatory, University of California, Santa Cruz,
	     CA 95064, USA\\
         \and
             Adnet Systems, Inc., NASA Goddard Space Flight Center,
             Code 671, Greenbelt, MD 20771, USA\\
             }
   \authorrunning{Rots, Bunclark, Calabretta, Allen, Manchester, \& Thompson}

   \date{Received 22 July 2014 / Accepted 26 September 2014}
% \abstract{}{}{}{}{} 
% 5 {} token are mandatory
 
  \abstract
   {In a series of three previous papers, formulation and specifics
    of the representation of World Coordinate Transformations in FITS data
    have been presented. This fourth paper deals with encoding
    time.}
   {Time on all scales and precisions known in astronomical datasets
    is to be described in an unambiguous, complete, and
    self-consistent manner.}
   {Employing the well--established World Coordinate System (WCS)
    framework, and maintaining compatibility with the FITS conventions
    that are currently in use to specify time, the standard is
    extended to describe rigorously the time coordinate.}
   {World coordinate functions are defined for temporal axes sampled
    linearly and as specified by a lookup table. The resulting
    standard is consistent with the existing FITS WCS standards and
    specifies a metadata set that achieves the aims enunciated above.}
   {}


   \maketitle
%
%________________________________________________________________

\normal

\section{Introduction}

Time as a dimension in astronomical data presents challenges in its
representation in FITS files as great as those met by the previous
papers in this series. The first, Paper I (Greisen \& Calabretta
2002), lays the groundwork by developing general constructs and
related FITS header keywords and the rules for their usage in
recording coordinate information.  Paper II (Calabretta \& Greisen
\cite{paper2}) addresses the specific problem of describing celestial
coordinates in a two-dimensional projection of the sky.  In Paper
III, Greisen et al.\ (\cite{paper3}) apply these methods to spectral
coordinates.  A draft paper (Calabretta et al.\ \cite{paper4})
proposes an extension to the formalism in order to deal with general
distortions of the coordinate grid.

This paper, the fourth in the series, formulates the representation 
of the time axis, or possibly multiple time axes, into the
FITS World Coordinate System (WCS) previously described.  We show how much
of the basic structure is employed, while developing extensions to
cope with the differences between time and other dimensions; notable
amongst these differences is the huge dynamic range, covering
the highest resolution timing relative to the age of the Universe.

The precision with which any time stamp conforms to any conventional
time scale is highly dependent on the characteristics of the acquiring
system.
The definitions of many conventional time scales vary over their
history along with the precision that can be attributed to any time
stamp.
The meaning of any time stamp may be ambiguous if a time scale is used
for dates prior to its definition by a recognized
authority, or for dates after that definition is abandoned.
However, common sense should prevail and it would be overly pedantic
to require a precision in the description of the time coordinate that
far exceeds the accuracy of the temporal information in the data.

In the following sections we first define the terms of reference
of this standard (Section \ref{refterms}).
Section \ref{treps} deals with time values and
representations of time. Section \ref{components} forms the core of
this standard, providing an explanation of the components that are
involved and defining the keywords to be used for specifying those
components.
Section \ref{implement} provides some general comments on implementing
this standard.
Section \ref{ucontext} on usage context refers back to the terms of
reference, illustrated with six header examples and including a
sub-section on time-related coordinate axes.
The prescriptive part of this standard is contained in Sections
\ref{refterms}, \ref{treps}, \ref{components},
and Appendix \ref{Appendix}.

Generally helpful references may be found in Seidelmann (\cite{seidel}),
Urban \& Seidelmann (\cite{urban}),
and McCarthy \& Seidelmann (\cite{McC+S}). The report on the current
(IAU 2009) system of astronomical constants is provided by Luzum
et al.\ (\cite{luzu})
\footnote{Current Best Estimates are maintained at\\
http://maia.usno.navy.mil/NSFA/NSFA\_cbe.html}.

The interpretation of the terms {\em must}, {\em required}, {\em
  should}, and {\em may} follows common usage in standards documents:\\
An implementation is compliant if it satisfies all the {\em must} or
{\em required} level requirements for the protocols it implements. An
implementation that satisfies all the {\em must} or {\em required}
level and all the {\em should} level requirements for its protocols is
said to be ``unconditionally compliant''; one that satisfies all the
{\em must} level requirements but not all the {\em should} level
requirements for its protocols is said to be ``conditionally
compliant''.
Alternatively, one may apply the common-sense compliance criterion
that requires all keywords needed for full interpretation of the data
to be present, with the exception of those for which default values
have been defined.

%__________________________________________________________________

\section{Terms of Reference}
\label{refterms}
Time WCS information needs to be supported in five contexts:
\begin{itemize}
 \item Recording time stamps in header keywords
 \item Time coordinate axes in images
 \item Time columns in tables
 \item Time coordinate axes in table vector columns
 \item Time in random groups\footnote{This structure is deprecated;
     we include it in this standard for completeness, but this should
     not be construed as a statement in support of its continued use.}
\end{itemize}
We distinguish the following components in the specification of time:
\begin{itemize}
 \item Time coordinate frame, containing:
 \begin{itemize}
  \item Time scale
  \item Reference time (the zero point for relative times)
  \item Time reference position
  \item Time reference direction (if applicable)
  \item Solar System ephemeris used (if applicable)
 \end{itemize}
 \item Time unit
 \item Corrections, errors, etc.:
 \begin{itemize}
  \item Time offsets
  \item Absolute error
  \item Relative error
  \item Time resolution
 \end{itemize}
 \item Durations
\end{itemize}

The following use cases illustrate the scope of the
requirements for time axes.
\begin{itemize}
  \item Photon arrival times (``event lists'')
  \item Time-sampled data streams (referred to as ``light curves'' in
  some of our communities)
  \item Pulsar pulse profiles and other folded or stacked light curves
  \item Power spectra, cross-, and auto-correlation spectra
  \item Image cubes: typically a series of two-dimensional images acquired
at regular time spacing, and stacked so the third axis is time. Usually
precision is not demanding, but the time axis must be integrated into a
three-dimensional WCS.
  \item Simulation data
\end{itemize}
``Mixed'' axes, where spatial or spectral coordinates change as a
function of time (e.g., during an observation) represent a special
challenge.

Where possible, we have adopted the same keywords as in the 
OGIP convention\footnote{This convention was developed by the 
Office of Guest Investigator Programs within the HEASARC (High Energy 
Astrophysics Science Archive Research Center) at the NASA Goddard Space 
Flight Center; (OGIP/GSFC/NASA~\cite{nasa}).},
which has become a {\em de facto} standard for representing 
timing information within high-energy astrophysics data files, 
particularly from NASA as well as many ESA missions.

In addition to
absolute time axes, we provide accommodation for
three types of time-related coordinates: Phase, Timelag, and
Frequency; see Section \ref{related}.

Contrary to the convention followed in previous FITS standards papers,
Appendix \ref{Appendix} is to be considered part of this standard. The
more subtle issues associated with the definition of time scales are,
of necessity, germaine to the details of the standard, but it seemed
unwieldy to include them in the main text of this paper.

\section{Time Values and Representations of Time}
\label{treps}
The three most common ways to specify time are:
ISO-8601, Julian Date (JD; see Herschel~\cite{herschel}),
or Modified Julian Date (MJD = JD
$-$ 2,400,000.5; see IAU~\cite{IAUmjd}).
Julian Dates are counted from Julian proleptic calendar date 1
January 4713 BCE at noon, or Gregorian proleptic calendar date 24
November 4714 BCE, also at noon. For an explanation of the calendars,
see the note in Section \ref{iso8601}.

Even though we may tend to think of certain representations of time as
absolute (ISO-8601, Julian dates), time values in this paper
{\em shall} all
be considered relative: elapsed time since a particular reference point
in time. It may help to view the ``absolute'' values
as merely relative to a globally accepted zero point.

In the following we first treat the ISO-8601 representation, then
floating point values of elapsed time since a reference value.
For time, more than any other coordinate, precision may be a concern
and naive use of double precision floating point parameters for
time values (especially Julian Dates) will be inadequate in some
cases. However, a judicious combination of keywords and their
values, as described in the remainder of this section, will allow
almost any required precision to be achieved without having to resort
to anything beyond double precision data types in handling keyword
values. We urge creators of data products to apply special care, so
that clients can rely on this being the case.
If and when, in addition to the 32-bit ({\tt E}) and 64-bit ({\tt D})
floating point types, a 128-bit floating point data type becomes
available and supported, we envision that such a type 
will also be used for time values, removing the need for any special
provisions.

We conclude the section with a specification of epochs.

\subsection{ISO-8601 {\it Datetime} Strings}
\label{iso8601}
FITS uses a subset of ISO-8601 (which in itself does not imply a
particular time scale) for several time-related keywords
(Bunclark \& Rots~\cite{y2k}), such as \keyw{DATE-xxx}. 
In this paper we will use {\it datetime} as a pseudo data type to
indicate its use. Following the current FITS standard (Pence et al.\ 
\cite{pence}) its values {\em must} be written as a character string in
\keyv{A} format, but if and when an ISO-8601 data type is adopted,
it {\em should} be used in table columns, rather than the string type.

The full specification for the format of the {\it datetime} string
till now has been:
\begin{quote}
\centering
\keyv{CCYY-MM-DD[Thh:mm:ss[.s...]]}
\end{quote}
All of the time part {\em may} be omitted (just leaving the date) or the
decimal seconds may be omitted. Leading zeroes {\em must not} be omitted
and timezone designators are not allowed.

This paper extends the definition to allow five-digit years with a
(mandatory) sign, in accordance with ISO-8601. I.e., one {\em shall} either
use the {\em unsigned} four-digit year format or the {\em signed}
five-digit year format:
\begin{quote}
\centering
\keyv{[$\pm$C]CCYY-MM-DD[Thh:mm:ss[.s...]]}
\end{quote}

Note the following:
\begin{itemize}
\item
    In counting years, ISO-8601 follows the convention established by
    Cassini (\cite{cassini}) of including year zero.  Consequently, for negative
    year numbers there is an offset of one from BCE dates which do not
    recognize a year zero.  Thus year 1 corresponds to 1 CE, year 0 to
    1 BCE, year $-1$ to 2 BCE, and so on.
\item
The earliest date that may be represented in the four-digit year format
is 0000-01-01T00:00:00 (in the year 1 BCE); the latest date is
9999-12-31T23:59:59. This representation of time is tied to the
Gregorian calendar (Pope Gregorius XIII~\cite{gregory})\footnote{
The Gregorian calendar was an improvement of the Julian calendar which
was introduced by Julius Caesar in 46 BCE (Plinius~\cite{plinius}),
advised by Sosigenes of Alexandria and probably based on an earlier
design by Ptolemaeus III Euergetes (246-221 BCE) (Encyclop{\ae}dia
Britannica~\cite{EB}). Although the Julian calendar took effect in 45 BCE,
the leap days were not properly implemented until at least 4 CE,
requiring proleptic use of this calendar before that time. The
Gregorian calendar reduced the average length of the year from 365.25
days (Julian calendar) to 365.2425 days, closer to the length of the
tropical year. Further resources on these subjects can also be found
in Wikipedia.}.
In conformance with the present ISO-8601:2004(E) standard
(ISO~\cite{iso}) we specify that, for use in FITS files, dates prior
to 1582 {\em must} be
interpreted according to the proleptic application of the rules of
Gregorius XIII (\cite{gregory}). For dates not covered by the range we
{\em recommend} the use of Modified Julian Date (MJD) or Julian Date (JD)
numbers or the use of the signed five-digit year format. 
\item
In the five-digit year format the earliest and latest dates are\\
$-$99999-01-01T00:00:00 (i.e., $-100000$ BCE) and\\ $+$99999-12-31T23:59:59.
\item
Recalling the definition of JD provided at the top of Section
\ref{treps}, we can express its origin as
$-$04713-11-24T12:00:00.
\item
In time scale UTC the integer part of the seconds field runs from 00
to 60 (in order to accommodate leap seconds); in all other time scales the range is 00 to 59.
\item
This data type is not allowed in image axis descriptions since \keyw{CRVAL}
is required to be a floating point value.
\item
ISO-8601 {\it datetime} does not imply the use of any particular time
scale (see Section~\ref{timescale}).
\item
As specified by Bunclark \& Rots (\cite{y2k}), time zones are
explicitly not supported in FITS and, consequently, appending the letter `Z'
to a FITS ISO-8601 string is not allowed. The rationale for this rule
is that its role in the ISO standard is that of a time zone indicator,
not a time scale indicator. As the concept of a time zone is not
supported in FITS, the use of time zone indicator is inappropriate.
\end{itemize}

\subsection{Single or Double Precision Floating Point Relative
  Time}
These are existing data types that do not need any particular
provisions and can be used when their precision suffices.
In general, if higher precision is required, it will be possible to
achieve this by judicious use of keyword values, without having
to resort to exotic datatypes, as described in the next subsection.

\subsection{Higher Precision in Keyword Values}
\label{longkeys}
While the FITS standard
(Pence, et al.\ \cite{pence}, Section 4.2.4). allows header
  values to be written to as many as 70 decimal digits, it must
  be recognised that practical implementations are currently
  based on double-precision floating point values which are
  capable of representing only approximately 15 decimal digits.
  While this has not been a limitation in the past, it may not
  be adequate for certain high-precision timing applications.
  In the absence of the widespread availability of quad-precision
  floating point, timing software often obtains the
  extra required precision by using a pair of double-precision
  values, typically containing the integer and fractional part,
  whose (implied) sum forms the high-precision value.  In like
  vein we provide the \keyw{[M]JDREF[IF]} and \keyw{DATEREF} keywords
  (see Section~\ref{trefval})
to define a global time reference epoch to which all times in the
  HDU are relative, and these should be used judiciously where
  high-precision timing is required.  Implementers of this
  standard should be aware that precision may be lost by adding
  relative times to the reference epoch, and maintain them
  as separate quantities until a final value is required (see
  Section~\ref{precision}).

\subsection{Higher Precision in Binary Tables: Doublet Vectors}
\label{doublets}
In binary tables one {\em may} use pairs of doubles. The time column in such a
table {\em shall} contain a vector of two doubles
(\TFORM{n} = \keyv{`2D'})
where the first component of
the doublet contains the integer portion of the time value and the
second one the fractional part,
such that their sum equals the true value and where both have the same sign.
This will ensure that retention of precision can be effected in as
simple a way as possible and avoiding any sign ambiguities.
We readily admit that a combination of
an integer and a floating point number would be preferable, but the
use of two doubles allows us to keep the time stamps in a single table
column.

\subsection{Julian and Besselian Epochs}
\label{epochs}
In a variety of contexts {\em epochs} are provided with astronomical
data. Until 1976 these were commonly based on the Besselian year (see
Section \ref{tunit}), with standard epochs B1900.0 and B1950.0. After
1976 the transition was made to Julian epochs based on the Julian year
of 365.25 days, with the standard epoch J2000.0. They are tied to time
scales ET and TDB, respectively. Table \ref{tab:epoch} provides
conversion values for some Besselian and Julian epochs. See also
Seidelmann (\cite{seidel}, Table 15.3).
Note that the Besselian epochs are scaled by the variable length
of the Besselian year (see Section \ref{tunit} and its cautionary
note, which also applies to this context).
The Julian epochs are easier to calculate, as long as one keeps
track of leap days.

See Section \ref{JBepoch} for the use of Besselian and Julian epochs
in FITS files.

\begin{table}
\caption{Some Besselian and Julian Epochs}
\label{tab:epoch}
\centering
\begin{tabular}{c l l}
\hline\hline
Epoch & ISO-8601 date & Julian Date \\
\hline
B1900 & 1899-12-31T19:31:26.4(ET) & 2415020.3135(ET)\\
B1950 & 1949-12-31T22:09:50.4(ET) & 2433282.4235(ET)\\
J1900 & 1899-12-31T12:00:00(ET)  &  2415020.0(ET)\\
J2000 & 2000-01-01T12:00:00(TDB) & 2451545.00(TDB)\\
J2001 & 2000-12-31T18:00:00(TDB) & 2451910.25(TDB)\\
J2002 & 2002-01-01T00:00:00(TDB) & 2452275.50(TDB)\\
J2003 & 2003-01-01T06:00:00(TDB) & 2452640.75(TDB)\\
J2004 & 2004-01-01T12:00:00(TDB) & 2453006.00(TDB)\\
\hline %inserts single line
\end{tabular}
\end{table}

Caution: be aware of the offset of 1 in negative year numbers,
compared with BCE dates (see Section \ref{iso8601}).

\section{Components of the Standard}
\label{components}
This section describes the components of the standard. The keywords
used to specify times are summarized in Table \ref{table:key}.
Section \ref{table:key}.a of the table contains data items: time values that have,
in principle, global validity in the HDU. Section \ref{table:key}.b presents
keywords that define the time reference frame for all time
values in the HDU and their context-dependent override keywords.
If the HDU contains a table, all keywords in the
first two sections {\em may} be replaced by columns, with specific values
for each row (``Green Bank convention''; Pence~\cite{pence2}).
The last section of the table (\ref{table:key}.c) lists the
keywords that allow overriding the global HDU keyword values for the
time axis in images.

In the following 
{\em datetime-valued} {\em must} be interpreted as
    {\em string-valued} where the string conforms to ISO-8601 format
    as defined in Section \ref{iso8601}
.

\subsection{Time Coordinate Frame}
This section defines the various components that constitute the time
coordinate frame.
For a full review of the IAU resolutions concerning space-time
coordinate transformations, see Soffel et al.\ (\cite{soffel}).

\subsubsection{Time Scale}
\label{timescale}

The time scale defines the temporal reference frame (in the
terminology of the IVOA Space-Time Coordinate metadata standard; see
Rots~\cite{rots}).  See also the USNO (\cite{usno}) page on time
scales, Wallace (\cite{wall}), and SOFA (\cite{sofa}).

Table \ref{table:timescale} lists recognized values. For a detailed
discussion of the time scales we refer to Appendix \ref{Appendix}; that
information will be of particular relevance for high-precision
timing. In cases where this 
is significant, one may append a specific realization, in parentheses,
to the values in the table; e.g., \keyv{TT(TAI)}, \keyv{TT(BIPM08)},
\keyv{UTC(NIST)}.
Note that linearity is not preserved across all time scales.
Specifically, if the reference position remains unchanged (see Section
\ref{trefpos}), the first ten, with the exception of \keyv{UT1}, are
linear transformations of each
other (excepting leap seconds), as are \keyv{TDB} and \keyv{TCB}.
On average \keyv{TCB} runs faster than \keyv{TCG} by approximately
$1.6 \times 10^{-8}$, but the transformation from \keyv{TT} or
\keyv{TCG} (which are linearly related) is to be achieved through a
time ephemeris, as provided by Irwin \& Fukushima 
(\cite{IF99}).

The relations between coordinate time scales and their dynamical
equivalents have been defined as (see Luzum et al.\ \cite{luzu},
Wallace~\cite{wall}, SOFA~\cite{sofa}):\\
{\indent $T(\mathrm{TCG}) = T(\mathrm{TT}) + L_{\mathrm{G}} \times
  86400 \times (JD(\mathrm{TT}) - JD_0)$}\\
{\indent $T(\mathrm{TDB}) = T(\mathrm{TCB}) - L_{\mathrm{B}} \times
  86400 \times (JD(\mathrm{TCB}) - JD_0) + TDB_0$}\\
where:\\
{\indent $T$ is in seconds}\\
{\indent $L_{\mathrm{G}} = 6.969290134 \times 10^{-10}$}\\
{\indent $L_{\mathrm{B}} = 1.550519768 \times 10^{-8}$}\\
{\indent $JD_0 = 2443144.5003725$}\\
{\indent $TDB_0 = -6.55 \times 10^{-5}$} s\\

Linearity is virtually guaranteed since images and individual table
columns do not allow more than one reference position to be associated
with them and since there is no overlap between reference positions that are
meaningful for the first nine time scales on the one hand, and for the
barycentric ones on the other.
All use of the time scale GMT in FITS files {\em shall} be taken to have its
zero point at midnight, conformant with UT, including dates prior to
1925; see Sadler (\cite{sadler}). For high-precision timing prior to
1972, see Section \ref{ut}.

\begin{table}
\caption{Recognized Time Scale Values$^{1,2}$}
\label{table:timescale}
\centering
\begin{tabular}{r p{0.35\textwidth}}
\hline\hline
\\
\keyw{TAI} & (International Atomic Time): atomic time standard
maintained on the rotating geoid\\
\keyw{TT} & (Terrestrial Time; IAU standard): defined on the rotating
geoid, usually derived as TAI $+$ 32.184 s\\
\keyw{TDT} & (Terrestrial Dynamical Time): synonym for TT (deprecated)\\
\keyw{ET} & (Ephemeris Time): continuous with TT; should not be used for data taken
after 1984-01-01\\
\keyw{IAT} & synonym for TAI (deprecated)\\
\keyw{UT1} & (Universal Time): Earth rotation time\\
\keyw{UTC} & (Universal Time, Coordinated; default): runs synchronously with TAI,
except for the occasional insertion of leap seconds intended to keep
UTC within 0.9 s of UT1;\\
 & as of 2012-07-01 UTC = TAI $-$ 35 s\\
\keyw{GMT} & (Greenwich Mean Time): continuous with UTC; its use is
deprecated for dates after 1972-01-01\\ 
\keyw{UT()}& (Universal Time, with qualifier): for high-precision use
of radio signal distributions between 1955 and 1972; see Section \ref{ut}\\
\keyw{GPS} & (Global Positioning System): runs (approximately) synchronously with TAI;
GPS $\approx$ TAI $-$ 19 s\\
\keyw{TCG} & (Geocentric Coordinate Time): TT reduced to the geocenter,
corrected for the relativistic effects of the Earth's rotation and
gravitational potential; TCG runs faster than TT at a constant
rate\\
\keyw{TCB} & (Barycentric Coordinate Time): derived from TCG by a
4-dimensional transformation, taking into account the relativistic
effects of the gravitational potential at the barycenter (relative
to that on the rotating geoid) as well as velocity time dilation
variations due to the eccentricity of the Earth's orbit,
thus ensuring consistency with fundamental physical constants; Irwin
\& Fukushima (\cite{IF99}) provide a time ephemeris\\
\keyw{TDB} & (Barycentric Dynamical Time): runs slower than TCB at a
constant rate so as to remain approximately in step with TT;
runs therefore quasi-synchronously with
TT, except for the relativistic effects introduced by variations
in the Earth's velocity relative to the barycenter; when referring
to celestial observations, a pathlength correction to the barycenter
may be needed which requires the Time Reference Direction used in
calculating the pathlength correction\\
\keyw{LOCAL} & for simulation data and for free-running clocks.\\
 & \\
\hline
\multicolumn{2}{l}{$^1$Specific realizations may be appended to these values,}\\
\multicolumn{2}{l}{in parentheses; see text. For a more detailed discussion of}\\
\multicolumn{2}{l}{time scales, see  Appendix \ref{Appendix}}\\
\multicolumn{2}{l}{$^2$Recognized values for \keyw{TIMESYS}, \CTYPE{ia}, \TCTYP{n}, \TCTY{na}.}
\end{tabular}
\end{table}

Global Navigation Satellite System (GNSS) time scales GLONASS,
Galileo, and Beidou are not included, as they are less mature and/or
less widely used than GPS. They may be added in the future if their
use becomes more common in the scientific community.

Other time scales that are not listed in Table
\ref{table:timescale} are intrinsically unreliable and/or ill-defined.
These {\em should} be tied to one of the existing scales with appropriate
specification of the uncertainties; the same is true for free-running
clocks. However, a local time scale, such as MET (Mission Elapsed
Time) or OET (Observation Elapsed Time), {\em may} be defined for practical
reasons. In those cases the time reference value (see Section \ref{trefval})
{\em shall not} be applied to the
values and we strongly {\em recommend} that such
timescales be provided as alternate time scales, with a defined
conversion to a recognized time scale.

Most current computer operating systems adhere to the POSIX standard
for time, and use Network Time Protocol (NTP) to synchronize closely
to UTC.  This reasonable approximation to UTC is then commonly used
to derive timestamps for FITS data. However, authors of FITS writers
and subsequent users of FITS timing information should be aware of
the accuracy limitations of POSIX and NTP, especially around the
time of a leap second.

Finally, it may be helpful, in order to put the different time scales
into perspective, to emphasize that while UT1 is, in essence, an angle
(of the Earth's rotation -- {\em i.e.,} a {\em clock}), the others are
SI-second counters ({\em chronometers}); UTC, by employing
leapseconds, serves as a bridge between the two types of time scales.

\paragraph{Keywords}
\label{ktimescale}

The global keyword that records the time scale is
\newkeyp{\keyw{TIMESYS}}{string}
{Time scale; default \keyv{UTC}}

In relevant context ({\em e.g.,} time axes in image arrays, table
columns, or random groups)
it {\em may} be overridden by a time scale recorded in \CTYPE{ia},
its binary table equivalents, or \PTYPE{i} (see Table \ref{table:key}).

The keywords \TIMESYS{}, \CTYPE{ia}, \TCTYP{n}, and \TCTY{na} {\em may}
assume the values listed in Table \ref{table:timescale}.
In addition, for backward compatibility, all except \TIMESYS{} and \PTYPE{i}
{\em may} also assume the value \keyv{TIME} (case-insensitive), whereupon
the time scale {\em shall} be that recorded in \TIMESYS{} or, in its absence, its default value, \keyv{UTC}.
See also Sections \ref{images}, \ref{timecol}, and \ref{random}.
See Section \ref{related} regarding their use for specific time-related axes.

As noted above, local time scales other than those listed in Table
\ref{table:timescale} {\em may} be used, but their use
{\em should} be restricted
to alternate coordinates in order that the primary coordinates will
always refer to a properly recognized time scale; an example may be
found in Section \ref{timecol}.

\subsubsection{Time Reference Value}
\label{trefval}
We allow the time reference point to be defined in the three common
systems: ISO-8601, JD, or MJD. These reference values are
only to be applied to time values associated with one of the
recognized time scales listed in Table \ref{table:timescale}
and that time scale needs to be specified (see also Section
\ref{labeling}).

\paragraph{Keywords}
\label{ktrefval}
The reference point in time, to which all times in the HDU are
relative, {\em shall} be specified through one of three keywords:
\newkeyp{\keyw{MJDREF}}{floating}
{Reference time in MJD}
\newkeyp{\keyw{JDREF}}{floating}
{Reference time in JD}
\newkeyp{\keyw{DATEREF}}{datetime}
{Reference time in ISO-8601}
\keyw{MJDREF} and \keyw{JDREF} {\em may}, for clarity and/ or precision
reasons, be split into two keywords holding the integer and fractional
parts separately:
\newkeyp{\keyw{MJDREFI}}{integer}
{Integer part of reference time in MJD}
\newkeyp{\keyw{MJDREFF}}{floating}
{Fractional part of reference time in MJD}
\newkeyp{\keyw{JDREFI}}{integer}
{Integer part of reference time in JD}
\newkeyp{\keyw{JDREFF}}{floating}
{Fractional part of reference time in JD}
If \keyw{[M]JDREF} and both \keyw{[M]JDREFI} and \keyw{[M]JDREFF} are
present, the integer and fractional values shall have precedence over
the single value. If the single value is present with one of the two
parts, the single value shall have precedence.
In the following, \keyw{MJDREF} and \keyw{JDREF} refer to their
literal meaning or the combination of their integer and fractional parts.

If, for whatever reason, a header
contains more than one of these keywords, \JDREF{} {\em shall} have
precedence over \DATEREF{} and \MJDREF{} {\em shall} have precedence over
both the others. If none of the three keywords is present, there is no
problem as long as all times in the HDU are expressed in ISO-8601;
otherwise \MJDREF{} = 0.0 {\em must} be assumed.
If \TREFPOS{} = '\keyv{CUSTOM}' (Section \ref{trefpos}) it is
legitimate for none of
the reference time keywords to be present, as one may assume that we
are dealing with simulation data.

\paragraph{Note:}
The {\em value} of the reference time has global validity for all
time values, but it does not have a particular time scale associated
with it. \\
Therefore, assuming the use of TT(TAI), if \MJDREF{} = 50814.0 
and \TIMEUNIT = '\keyv{s}':\\
 a time instant $T = 86400.0$
 associated with TT will fall on\\
  {\indent 1998-01-02T00:00:00.0(TT) or}\\
  {\indent 1998-01-01T23:59:27.816(TAI) or}\\
  {\indent 1998-01-01T23:58:56.816(UTC),}\\
 but a time instant $T = 86400.0$
 associated with TAI will fall on\\
  {\indent 1998-01-02T00:00:32.184(TT) or}\\
  {\indent 1998-01-02T00:00:00.0(TAI) or}\\
  {\indent 1998-01-01T23:59:29.0(UTC).}\\
Table \ref{tab:example4} provides examples of this; one may compare
the reference pixel values of TT, TCG, and UTC for column 1, and of
TDB and TCB for column 20.

\subsubsection{Time Reference Position}
\label{trefpos}

An observation is an event in space-time.
The reference position, specified by the keyword \keyw{TREFPOS},
specifies the spatial location at which the
time is valid,
either where the observation was made or the point in space for which
light-time corrections have been applied.
This {\em may} be a standard location (such as \keyv{GEOCENTER}
or \keyv{TOPOCENTER}) or a point in space defined by specific coordinates.
In the latter case one should be aware that a (3-D) spatial coordinate
frame needs to be defined that is likely to be different from the
frame(s) that the data are associated with. Note that \keyv{TOPOCENTER}
is only moderately informative if no observatory location is provided
or indicated.

The commonly allowed standard values are shown in Table~\ref{table:refpos}.
Note that for the gaseous planets we use the barycenters of their
planetary systems, including satellites, for obvious reasons.
Our preference is to spell the location names out in full, but in
order to be consistent with the practice of Paper III (\cite{paper3})
and the FITS Standard (Pence, et al.\ \cite{pence}) the values are
allowed to be truncated to eight characters. 
Furthermore, in order to allow for alternative spellings, only the
first three characters of all these values {\em shall} be considered
significant. The value of the keyword {\em shall be} case-sensitive.
We envisage that at some
time in the future we may need a provision to add minor planets to
this list.

\begin{table}
\caption{Standard Time Reference Position Values$^1$}
\label{table:refpos}
\centering
\begin{tabular}{r p{0.30\textwidth}}
\hline\hline
\\
\keyw{TOPOCENTER} & Topocenter: the location from where the
observation was made (default)\\
\keyw{GEOCENTER} & Geocenter\\
\keyw{BARYCENTER} & Barycenter of the Solar System\\
\keyw{RELOCATABLE} & Relocatable: to be used for simulation data only\\
\keyw{CUSTOM}   & A position specified by coordinates that is not the
observatory location\\
\hline
 \noalign{\vskip 1ex}% 
\multicolumn{2}{l}{Less common allowed standard values are:}\\
 \noalign{\vskip .8ex}% 
\hline
\keyw{HELIOCENTER} & Heliocenter\\
\keyw{GALACTIC} & Galactic center\\
\keyw{EMBARYCENTER} & Earth-Moon barycenter\\
\keyw{MERCURY}  & Center of Mercury\\
\keyw{VENUS}    & Center of Venus\\
\keyw{MARS}     & Center of Mars\\
\keyw{JUPITER}  & Barycenter of the Jupiter system\\
\keyw{SATURN}   & Barycenter of the Saturn system\\
\keyw{URANUS}   & Barycenter of the Uranus system\\
\keyw{NEPTUNE}  & Barycenter of the Neptune system\\
\\
\hline
\multicolumn{2}{l}{$^1$Recognized values for \keyw{TREFPOS},
  \TRPOS{n};}\\
\multicolumn{2}{l}{only the first three characters of the
    values are significant and}\\
\multicolumn{2}{l}{solar system locations are as specified in the JPL Ephemerides}
\end{tabular}
\end{table}

Some caution is in order here. Time scales and reference positions
cannot be combined arbitrarily if one wants a clock that runs linearly
at \keyw{TREFPOS}. Table \ref{table:posscale} provides a
summary of compatible combinations.
\keyv{BARYCENTER} {\em should} only be used in conjunction with time scales
\keyv{TDB} and \keyv{TCB} and {\em should} be the only reference position
used with these time scales.
With proper care
\keyv{GEOCENTER}, \keyv{TOPOCENTER}, and \keyv{EMBARYCENTER} are
appropriate for the first ten time scales in 
Table \ref{table:timescale}. However, one needs to be aware that
relativistic effects introduce a (generally linear) scaling in certain
combinations; highly eccentric spacecraft orbits
are the exceptions. 
Problems will arise when using a reference position on another solar
system body (including \keyv{HELIOCENTER}). At this point we {\em recommend}
synchronizing the local clock with one of the time scales defined on
the Earth's surface, \keyv{TT}, \keyv{TAI}, \keyv{GPS}, or \keyv{UTC}
(in the last case: beware of leap seconds). This is common practice
for spacecraft clocks. Locally, such a clock will
not appear to run at a constant rate, because of variations in the
gravitational potential and in motions with respect to Earth, but the
effects can be calculated and are probably small compared with errors
introduced by the alternative: establishing a local time standard.

\begin{table}
\caption{Compatibility of Time Scales and Reference Positions$^{1}$}
\label{table:posscale}
\centering
\begin{tabular}{l c c c c c}
\hline\hline
Reference & TT, TDT & TCG & TDB & TCB & LOCAL\\
Position & TAI, IAT &\\
 & GPS &\\
 & UTC, GMT &\\
\hline
\\
\keyv{TOPOCENTER} & t & ls & & & \\
\keyv{GEOCENTER} & ls & c & & & \\
\keyv{BARYCENTER} & & & ls & c & \\
\keyv{RELOCATABLE} & & & & & c\\
Other$^{2}$  & re & re & & & \\
\\
\hline
\multicolumn{6}{l}{$^1$Legend (combination is not recommended if no entry):}\\
\multicolumn{6}{l}{c: correct match; reference position coincides with the spatial origin}\\
\multicolumn{6}{l}{of the space-time coordinates}\\
\multicolumn{6}{l}{t: correct match on Earth's surface, otherwise usually linear scaling}\\
\multicolumn{6}{l}{ls: linear relativistic scaling}\\
\multicolumn{6}{l}{re: non-linear relativistic scaling}\\
\multicolumn{6}{l}{$^2$All other locations in the solar system}\\
\end{tabular}
\end{table}

In order to provide a complete description, \keyv{TOPOCENTER}
requires the observatory's coordinates to be specified.
We offer three options: the ITRS Cartesian coordinates (X, Y, Z)
introduced in Paper III; a geodetic latitude/longitude/height triplet;
or a reference to an orbit ephemeris file. 

A non-standard location indicated by \keyv{CUSTOM} {\em must} be specified
in a manner similar to the specification of the observatory location
(indicated by \keyv{TOPOCENTER}). One should be careful with the use
of the \keyv{CUSTOM} value and not confuse it with \keyv{TOPOCENTER},
as use of the latter imparts additional information on the provenance
of the data.

\paragraph{Keywords}
\label{ktrefpos}
The time reference position is specified by the keyword
\newkeyp{\keyw{TREFPOS}}{string}
{Time reference position; default \keyv{TOPOCENTER}}
\TREFPOS{}\footnote{The OGIP convention
uses the keyword \keyw{TIMEREF} and only allows values '\keyv{LOCAL}'
(i.e., Topocenter), '\keyv{GEOCENTRIC}', '\keyv{HELIOCENTRIC}',
'\keyv{SOLARSYSTEM}' (i.e., Barycenter); the convention contains
also the somewhat peculiar keyword \keyw{TASSIGN}. We will
not adopt these keywords in order to avoid confusion on allowed values
and meaning. Instead, we adopt the keywords \keyw{TREFPOS} and
\keyi{TRPOS}{n}.}
 {\em shall} apply to time coordinate axes in images as well. See
Section \ref{restrictions} for an explanation.

In binary tables different columns {\em may} represent completely different
Time Coordinate Frames. However, each column can have only one time
reference position, thus guaranteeing linearity (see Section \ref{timescale})
and the following keyword may override \TREFPOS{}:
\newkey{\TRPOS{n}}{string}
If the value of any of these keywords is \keyv{TOPOCENTER}, the
observatory position needs to be specified. If the value is
\keyv{CUSTOM}, the ``custom'' position needs to be specified.
In either case we allow three mechanisms for this.

The ITRS Cartesian coordinates (with respect to the geocenter) as
defined in Paper III:
\newkeyp{\keyw{OBSGEO-X}}{floating}
{ITRS Cartesian X in m}
\newkeyp{\keyw{OBSGEO-Y}}{floating}
{ITRS Cartesian Y in m}
\newkeyp{\keyw{OBSGEO-Z}}{floating}
{ITRS Cartesian Z in m}
Similarly defined geodetic coordinates have to be recognized, although
the ITRS Cartesian coordinates are strongly preferred:
\newkeyp{\keyw{OBSGEO-B}}{floating}
{Latitude in deg, North positive}
\newkeyp{\keyw{OBSGEO-L}}{floating}
{Longitude in deg, East positive}
\newkeyp{\keyw{OBSGEO-H}}{floating}
{Altitude in m}
An orbit ephemeris file:
\newkeyp{\keyw{OBSORBIT}}{string}
{URI, URL, or name of orbit ephemeris file}

Beware that only one set of coordinates is allowed in a given HDU.
Cartesian ITRS coordinates are the preferred coordinate system;
however, when using these in an environment requiring nanosecond
accuracy, one should take care to distinguish between meters
consistent with TCG or with TT.
If one uses geodetic coordinates, the geodetic altitude
\keyw{OBSGEO-H} is measured with respect to IAU
1976 ellipsoid which is defined as having a semi-major axis of
6378140~m and an inverse flattening of 298.2577. 

ITRS coordinates (X,Y,Z) may be derived from geodetic coordinates
(L,B,H) through:
  \[\indent X=(N(B)+H) \cos (L) \cos (B)\]
  \[\indent Y=(N(B)+H) \sin (L) \cos (B)\]
  \[\indent Z=(N(B) (1-e^2)+H) \sin (B)\]
where:
  \[\indent N(B)=\frac{a}{\sqrt{1-e^2 \sin^2(B)}}\]
  \[\indent e^2=2f - f^2\]
$a$ is the semi major axis, $f$ the inverse of the inverse flattening.

Nanosecond precision in timing requires that
\keyw{OBSGEO-[BLH]} be expressed in a geodetic reference frame defined
after 1984 in order to be sufficiently accurate.

\subsubsection{Time Reference Direction}
\label{trefdir}

If any pathlength corrections have been applied to the time stamps
(i.e., if the reference position is not \keyv{TOPOCENTER} for observational
data), the reference direction that is used in calculating the
pathlength delay {\em should} be provided in order to maintain a proper
analysis trail of the data. However, this is useful only
if there is also information available on the location from where the
observation was made (the observatory location). The direction will
usually be provided in a spatial coordinate
frame that is already being used for the spatial metadata, although
that is not necessarily the case. It is,
for instance, quite conceivable that multiple spatial frames are
already involved: spherical ICRS coordinates for celestial positions,
and Cartesian FK5 for spacecraft ephemeris.
We also acknowledge that the time reference direction does not by
itself provide sufficient information to perform a fully correct
transformation; however, within the context of a specific analysis
environment it should suffice.

The uncertainty in the reference direction affects the errors in the
time stamps. A typical example is provided by barycentric corrections
where the time error $t_{err}$ is related to the position error $pos_{err}$:
\[t_{err} \mbox{(ms)} \leq 2.4 pos_{err} \mbox{(arcsec)}\]

The reference direction is indicated through a reference to
specific keywords.
These keywords {\em may} hold the reference direction
explicitly or indicate columns holding the coordinates.
In event lists where the individual photons are tagged with a spatial
position, those coordinates {\em may} have been used for the reference
direction and the reference will point to the columns containing these
coordinate values.
The OGIP convention, on the other hand, uses the keywords
\keyw{RA\_NOM} and \keyw{DEC\_NOM} indicating a globally applied
direction for the entire HDU.

\paragraph{Keywords}
\label{ktrefdir}
The time reference direction is specified by the keyword
\newkeyp{\keyw{TREFDIR}}{string}
{Pointer to time reference direction}
\TREFDIR{} {\em shall} apply to time coordinate axes in images as well. See
Section \ref{restrictions} for an explanation.

In binary tables different columns {\em may} represent completely different
Time Coordinate Frames. However, also in that situation the condition
holds that each column can have only one Time Reference Direction.
Hence, the following keyword may override \TREFDIR{}:
\newkey{\TRDIR{n}}{string}
The value of the keyword {\em shall} consist of the name of the keyword or
column containing the longitudinal coordinate, followed by a comma,
followed by the name of the keyword or column containing the
latitudinal coordinate. For the above quoted OGIP convention this
would result in:\\
\indent \TREFDIR = \keyv{'RA\_NOM,DEC\_NOM'}\\
For the example in Table \ref{tab:example4}:\\
\indent \keyw{TRDIR20} = \keyv{'EventRA,EventDEC'}

\subsubsection{Solar System Ephemeris}
\label{plephem}

If applicable, the Solar System ephemeris used for
calculating pathlength delays {\em should} be identified. This is
particularly pertinent when the time scale is
\keyv{TCB} or
\keyv{TDB}.

The ephemerides that are currently most often used are JPL's (NASA/JPL~\cite{jple}
and \cite{jpl}):
\begin{itemize}
  \item \keyv{DE200} (Standish~\cite{ems1}; considered obsolete, but still in use)
  \item \keyv{DE405} (Standish~\cite{ems2}; default)
  \item \keyv{DE421} (Folkner et al.\ \cite{wmf})
  \item \keyv{DE430}, \keyv{DE431}, \keyv{DE432} (Folkner et al.\ \cite{wmfpk})
\end{itemize}
Future ephemerides in this series {\em shall} be accepted and recognized as they are released.

\paragraph{Keyword}
The Solar System ephemeris used for the data (if
required) is indicated by the value of the keyword
\newkeyp{\keyw{PLEPHEM}}{string}
{Solar System ephemeris; default \keyv{DE405}}
Historically, the name \keyw{PLEPHEM} referred to Planetary and Lunar
Ephemeris; we continue the use of that keyword name.

\subsection{Time Unit}
\label{tunit}

The specification of the time unit
allows the values defined in Paper I (\cite{paper1}) and the FITS
Standard (Pence, et al.\ \cite{pence}), with the
addition of the century. We {\em recommend} the following:
\begin{itemize}
  \item \keyv{s}: second (default)
  \item \keyv{d}: day (= 86,400 s)
  \item \keyv{a}: (Julian) year (= 365.25 d)
  \item \keyv{cy}: (Julian) century (= 100 a)
\end{itemize}
The following values are also acceptable:
\begin{itemize}
  \item \keyv{min}: minute (= 60 s)
  \item \keyv{h}: hour (= 3600 s)
  \item \keyv{yr}: (Julian) year (= \keyv{a} = 365.25 d)
  \item \keyv{ta}: tropical year
  \item \keyv{Ba}: Besselian year
\end{itemize}
The use of \keyv{ta} and \keyv{Ba} is not encouraged, but there are
data and applications that require the use of tropical years or
Besselian epochs (see Section \ref{epochs}).
The length of the tropical year \keyv{ta} in days is (based on Simon,
et al.\ \cite{simon}):\\
\indent \mbox{$ 1 \hspace{4 pt}\mathrm{ta} = 365.24219040211236 - 0.00000615251349\hspace{2 pt} T$}
\indent \mbox{$\hspace{30 pt} - 6.0921 \times 10{^{-10}}\hspace{2 pt} T{^2} + 2.6525 \times
  10{^{-10}}\hspace{2 pt} T{^3}\hspace{4 pt} \mathrm{d}$}\\ 
where $T$ is in Julian centuries since J2000, using time scale TDB.\\
The length of the Besselian year \keyv{Ba} in days is
(based on Newcomb~\cite{newc95} and \cite{newc98}):\\
\indent \mbox{$ 1 \hspace{4 pt} \mathrm{Ba} = 365.2421987817 -
  0.00000785423\hspace{2 pt} T \hspace{4 pt} \mathrm{d}$}\\ 
where $T$ is in Julian centuries since J1900, using time scale ET --
although for these purposes the difference with TDB is negligible.

A cautionary note is in order here. The subject of tropical and
Besselian years presents a particular quandary for the specification
of standards. The expressions presented here specify how to calculate
them for use in data files while creating these. However, that is
pretty much a non-statement since such practice is strongly
discouraged. Our purpose in providing the expressions is to guide the
user in how to interpret existing data that are based on
these units. But there is no guarantee that the authors of the data
applied these particular definitions and there is ample evidence that
many did not (see, {\em e.g.,} Meeus \& Savoie~\cite{meeus}). 
An alternative definition of the Besselian epoch in common use ({\em e.g.,}
in SOFA~\cite{sofa}) is the
one given by Lieske~\cite{lies}:
\indent \mbox{$ B = 1900.0 + (JD - 2415020.31352) / 365.242198781$}\\
which is based on a Besselian year of fixed length leading to:\\
\indent \mbox{$ 1 \hspace{4 pt} \mathrm{Ba} = 365.242198781 \mathrm{d}$}\\ 
Therefore,
all we can state here is that these are the most accurate available
expressions for the units, but at the same time we strongly advise any
user of existing data that contain them to pay special attention and
attempt to ascertain what the data's authors really used. 

See Section \ref{JBepoch} for the use of Besselian and Julian epochs
in FITS files.

\paragraph{Keywords}
\label{ktunit}
The time unit is set by the keyword 
\newkeyp{\keyw{TIMEUNIT}}{string}
{Time unit; default \keyv{s}}
that {\em shall} apply to all time instances and durations that do
  not have an implied time unit (such as is the case for JD, MJD,
  ISO-8601, J and B epochs).
In relevant context, this {\em may} be overridden (see Section
\ref{ucontext} for details)
by  the \CUNIT{ia} keywords and their
binary table equivalents (see Table \ref{table:key}).

\subsection{Assorted Items Affecting Time Data: Corrections, Errors, etc.}
\label{items}

All quantities enumerated below {\em must} be expressed in the prevailing
time units (\keyw{TIMEUNIT} or its local overrides), the default being \keyv{s}.

\subsubsection{Time Offset (not applicable to images)}
\label{tzero}

It is sometimes convenient to be able to apply a uniform clock
correction in bulk by just putting that number in a single keyword.
A second use for a time offset is to set a zero offset to a relative
time series, allowing zero-relative times, or just higher precision,
in the time stamps. Its default value is zero.

Its value {\em shall} be added to \keyw{MJDREF}, \keyw{JDREF}, or
\keyw{DATEREF}, and hence affects the values of \keyw{TSTART}, and
\keyw{TSTOP},
as well as any time pixel values in a binary table.

However, this construct {\em may} only be used in tables and {\em
  must not} be used in images.

\paragraph{Keyword}
\label{ktzero}
The time offset is set, in the units of \TIMEUNIT{}, by:
\newkeyp{\keyw{TIMEOFFS}}{floating}
{Time offset; default 0.0}
and has global validity for all times in the HDU. It has the same
meaning as the keyword \keyw{TIMEZERO} in the OGIP convention -- which
we did not adopt out of concern for the potentially ambiguous meaning
of the name. The net effect of this keyword is that the value of
\keyw{TIMEOFFS} is to be added to the time stamp values in the file.
Formally, this is effected by adding that value to \keyw{MJDREF},
\keyw{JDREF}, and/or \keyw{DATEREF}.

\subsubsection{Absolute Error}
\label{abserror}

The absolute time error is the equivalent of the systematic error
defined in previous papers.

\paragraph{Keywords}
\label{kabserror}
The absolute time error is set, in the units of \TIMEUNIT{}, by:
\newkeyp{\keyw{TIMSYER}}{floating}
{Absolute time error}
but {\em may} be overridden, in appropriate context
({\em e.g.,} time axes in image arrays or table columns;
see Section \ref{ucontext} for details)
by  the \CSYER{ia} keywords and their
binary table equivalents (see Table \ref{table:key}).

\subsubsection{Relative Error}
\label{relerror}
The relative time error specifies accuracy of the time stamps
relative to each other. This error will usually be much smaller
than the absolute time error.
This error is equivalent to the random error defined in previous papers.

\paragraph{Keywords}
\label{krelerror}
The relative time error (the random error between time stamps) is set,
in the units of \TIMEUNIT{}, by:
\newkeyp{\keyw{TIMRDER}}{floating}
{Relative time error}
but {\em may} be overridden, in appropriate context
({\em e.g.,} time axes in image arrays or table columns;
see Section \ref{ucontext} for details)
by  the \CRDER{ia} keywords and their
binary table equivalents (see Table \ref{table:key}).

\subsubsection{Time Resolution}
\label{timedel}

The resolution of the time stamps (the width of the time
  sampling function) is represented by a simple double.
In tables this may, for instance, be the size of the bins for time series data
or the bit precision of the time stamp values.

\paragraph{Keyword}
\label{ktimedel}
The time resolution is global in the HDU, and set by the keyword
\newkeyp{\keyw{TIMEDEL}}{floating}
{Time resolution}
in the units of \TIMEUNIT{}.

\subsubsection{Time Binning (not applicable to images)}
\label{timepixr}

When data are binned in time bins (or, as a special case, events
are tagged with a time stamp of finite precision) it is important
to know to which position in the bin (or pixel) that time stamp
refers. This is an important issue: the FITS standard
assumes that coordinate values correspond to the center of all pixels;
yet, clock readings are effectively truncations, not rounded values,
and therefore correspond to the lower bound of the pixel.

However, this construct {\em may} only be used in tables and {\em
  must not} be used in images.

\paragraph{Keyword}
\label{ktimepixr}
The relative position of the time stamp in each time bin (\TIMEDEL{} in
the case of an event list)
is set universally in the HDU by the keyword:
\newkeyp{\keyw{TIMEPIXR}}{floating}
{Pixel position of the time stamp; from 0.0 to 1.0, default 0.5.}
In conformance with the FITS pixel definition, the
default is 0.5, although the value 0.0 may be more common in certain
contexts. Note, for instance, that this is required when truncated
clock readings are recorded, as is the case for almost all event
lists. It seems unwise to allow this keyword
to be specified separately for multiple time frames, rather than
requiring its value to apply to all.

\subsection{Keywords that Represent Global Time Values}
\label{ktvals}

\paragraph{Keywords}
The following time values may only be found in the header, independent
of any time axes in the data. Except for \keyw{DATE}, they provide the
top-level temporal bounds of the data in the HDU. As noted before,
they may also be implemented as table columns.
\newkeyp{\keyw{DATE}}{datetime}
{Creation date of the HDU in UTC}
\newkeyp{\keyw{DATE-OBS}}{datetime}
{Time of data in ISO-8601 according to \keyw{TIMESYS}}
\newkeyp{\keyw{MJD-OBS}}{floating}
{Time of data in MJD according to \keyw{TIMESYS}}
\keyw{DATE-OBS} is already defined in Section 4.4.2.2 of the FITS
Standard. It is not specifically defined as the
start time of the observation and has also been used to indicate some form
of mean observing date and time. In order to specify a start date and
time unambiguously one {\em should} use:
\newkeyp{\keyw{DATE-BEG}}{datetime}
{Start time of data in ISO-8601 according to \keyw{TIMESYS}}
\newkeyp{\keyw{DATE-AVG}}{datetime}
{Average time of data in ISO-8601 according to \keyw{TIMESYS}; note: this
  paper does not prescribe how average times should be calculated}
\newkeyp{\keyw{DATE-END}}{datetime}
{Stop time of data in ISO-8601 according to \keyw{TIMESYS}}
\newkeyp{\keyw{MJD-BEG}}{floating}
{Start time of data in MJD according to \keyw{TIMESYS}}
\newkeyp{\keyw{MJD-AVG}}{floating}
{Average time of data in MJD according to \keyw{TIMESYS}; note: this
  standard does not prescribe how average times should be calculated}
\newkeyp{\keyw{MJD-END}}{floating}
{Stop time of data in MJD according to \keyw{TIMESYS}}
\newkeyp{\keyw{TSTART}}{floating}
{Start time of data in units of \keyw{TIMEUNIT} relative to \keyw{MJDREF},
\keyw{JDREF}, or \keyw{DATEREF} according to \keyw{TIMESYS}}
\newkeyp{\keyw{TSTOP}}{floating}
{Stop time of data in units of \keyw{TIMEUNIT} relative to \keyw{MJDREF},
\keyw{JDREF}, or \keyw{DATEREF} according to \keyw{TIMESYS}}
The alternate-axis equivalent keywords \keyi{DOBS}{n},
\keyi{MJDOB}{n},  \keyi{DAVG}{n}, and \keyi{MJDA}{n}, as defined in
the FITS Standard (Pence, et al.\ \cite{pence}, Table 22) are also allowed.
Note that of the above only \keyw{TSTART} and \keyw{TSTOP} are
relative to the time reference value.

As in the case of the time reference value (see Section \ref{trefval}), the
JD values supersede DATE values, and MJD values supersede both, in
cases where conflicting values are present.

It should be noted that, although they do not represent global time
values within an HDU, the \CRVAL{ia} and \CDELT{ia} keywords, and their binary
table equivalents (see Table \ref{table:key}), also represent (binary)
time values. They should be handled with the same care regarding
precision when combining them with the time reference value as any
other time value (see also Section \ref{precision}).

Finally, Julian and Besselian epochs (see Sections \ref{epochs} and
\ref{tunit}) {\em may} be expressed by these two keywords -- to be used with
great caution, as their definitions are more complicated and hence
their use more prone to confusion:
\newkeyp{\keyw{JEPOCH}}{floating}{Julian epoch; implied time scale TDB}
\newkeyp{\keyw{BEPOCH}}{floating}{Besselian epoch; implied time scale ET}
When these epochs are used as time stamps in a table column their
interpretation will be clear from the context. When the keywords
appear in the header without obvious context, they {\em must} be regarded
as equivalents of \keyw{DATE-OBS} and \keyw{MJD-OBS}, i.e., with no
fixed definition as to what part of the dataset they refer to.

\subsection{Other Time-related Coordinate Axes}
\label{related}
There are a few coordinate axes that are related to time and that are
accommodated in this standard: (temporal) {\em phase}, {\em timelag},
and {\em frequency}. {\em Phase}
results from folding a time series on a given period. {\em Timelag} is the
coordinate of cross- and auto-correlation spectra. As a practical
definition one may consider {\em frequency} as the Fourier transform
equivalent of time and, particularly, the coordinate axis of power 
spectra,
with the exception of spectra where the dependent variable is the
electromagnetic field. Specifically, the latter (excluded) case
applies to electromagnetic waveforms of cosmic origin with fixed
transformations to related variables such as wavelength, Doppler
velocity, and redshift which do not apply to periodic phenomena in
general.
That specific case is covered by Greisen et al.\ (\cite{paper3}).

These coordinate axes {\em shall} be specified by giving
\CTYPE{i} and its binary table equivalents one of the values:

\indent \keyv{PHASE}, \keyv{TIMELAG}, \keyv{FREQUENCY}\\
Note that the frequency coordinate of the electromagnetic spectrum
is indicated by the value \keyv{FREQ}.

{\em Timelag}'s units are the regular time units and {\em frequency's} basic unit is
\keyv{Hz}. Neither of these two coordinates is a linear or scaled
transformation of {\em time} and therefore cannot appear in parallel with
{\em time}  as an alternate description. Phrased differently,
a given vector of values for an observable
can be paired with a coordinate vector of {\em time}, or {\em
  timelag}, or {\em frequency}, but not 
with more than one of these; the three coordinates are orthogonal.

{\em Phase}, on the other hand, can appear in parallel with {\em time}
as an alternate description of the same axis.
Its units {\em may} be \keyv{deg}, \keyv{rad}, or \keyv{turn}, the
last of which is introduced here.

Time at the zero point of a {\em phase} axis {\em shall} be recorded in a new
keyword
      \newkey{\CZPHS{ia}}{floating}
with binary table forms \TCZPH{n}, \TCZP{na}, \iCZPH{i}{n}, and \iCZP{i}{na}.

Optionally, the period of a {\em phase} axis {\em may} be recorded in a new
keyword
      \newkey{\CPERI{ia}}{floating}
with binary table forms \TCPER{n}, \TCPR{na}, \iCPER{i}{n}, and \iCPR{i}{na}.
One should be aware, however, that this can be used only if the period
is a constant.
When that is not the case, the period {\em should} either be absent or set
to zero, and one should follow a convention like
PSRFITS\footnote{http://www.atnf.csiro.au/research/pulsar/index.html?n=Main.Psrfits}
(see also Hotan et al.\ \cite{hotan}, and Hobbs et al.\ \cite{hobbs}).

{\em Phase} period and zero point {\em shall} be expressed in the globally
valid time reference frame and unit as defined by the global keywords
(or their defaults) in the header.

\subsection{Durations}
\label{durations}

Durations {\em shall not} be expressed in ISO-8601 format, but only as
actual durations (i.e., numerical values) in the units of the
specified time unit.

There is an extensive collection of header keywords that indicate
time durations, such as exposure times, but there are many pitfalls
and subtleties that make this seemingly simple concept treacherous.
One may encounter similar-sounding keywords for concepts like: awarded
exposure time; scheduled exposure time; on-target time; duration of
the exposure, including dead time and lost time; exposure time charged
against the awarded exposure time; exposure time corrected for lost
(bad) data; and exposure time corrected for dead time.
Related to these are various keywords providing dead time correction
factors, dead time correction flags, and duty cycle information.
We suggest that these are
are excellent candidates for definition through an appropriate
formally registered FITS convention, rather than inclusion in
this standard.

Because of their crucial role and common use, keywords are defined here
to record exposure and elapsed time; in addition, a standard
for good time intervals is defined in Section \ref{GTI}.

\paragraph{Keyword}
\label{kdurations}
The only defined durations {\em shall} indicated by the keywords:
\newkey{\keyw{XPOSURE}}{floating}
in the units of \TIMEUNIT{}.
It {\em shall} be the effective exposure time for the data, corrected for
dead time and lost time. If the HDU contains multiple time slices, it
{\em shall} be the total accumulated exposure time over all slices.
More obvious candidates for the keyword name
(like \keyw{EXPOSURE}) had to be avoided since they have been used
with conflicting definitions in various sub-communities.
\newkey{\keyw{TELAPSE}}{floating}
also in the units of \TIMEUNIT{} provides the amount of time elapsed
between the start (\keyw{TSTART}, \keyw{MJD-BEG}, etc.) and the end
(\keyw{TSTOP}, \keyw{DATE-END}, etc.) of the observation or data stream.

\subsection{Good Time Interval (GTI) Tables}
\label{GTI}
Good-Time-Interval (GTI) tables are indispensable for data with
``holes'' in them, especially photon event files, as they allow one to
discriminate between ``no data received'' {\em versus} ``no data taken''.
GTI tables {\em shall} contain two mandatory columns, \keyw{START} and
\keyw{STOP}, and {\em may} contain one optional column, \keyw{WEIGHT}. The first two
define the interval, the third, with a value from 0 to 1, the quality
of the interval; {\em i.e.,} a weight of 0 indicates a {\em Bad}-Time-Interval.
\keyw{WEIGHT} has a default value of 1. Any time interval not covered
in the table shall be considered to have a weight of zero.

\section{General Comments on Implementation}
\label{implement}

In the following we discuss some practical implementation issues,
before turning, in the next section, to usage in specific contexts.

As a general comment, we should point out that the distortion
conventions described by Calabretta et al.\ (\cite{paper4}) are also
very much applicable to the time coordinate axis.

\subsection{Getting Started}
\label{getstarted}
As a simple getting-started guide, we make the following
recommendations (referring to Table \ref{table:key}):
\begin{itemize}
  \item The presence of the Informational \keyw{DATE} keyword is
    STRONGLY RECOMMENDED in all HDUs.
  \item One or more of the Informational keywords \keyw{DATE-xxx} and/or
    \keyw{MJD-xxx} SHOULD be present in all HDUs whenever a
      meaningful value can be determined. This also applies, for
      instance, to catalogs derived from data collected over a
      well-defined time range.
  \item The Global keyword \keyw{TIMESYS} is STRONGLY RECOMMENDED.
  \item The Global keywords \keyw{MJDREF} or \keyw{JDREF}
    or \keyw{DATEREF} are RECOMMENDED.
  \item The remaining Informational and Global keywords SHOULD be
    present whenever applicable.
  \item All Context-Specific keywords SHALL be present as needed and
    required by the context of the data.
\end{itemize}

\subsection{Global Keywords and Overrides}
\label{global}
For reference to the keywords that are discussed here, see
Table \ref{table:key}.
The globally applicable keywords listed in section \ref{table:key}.b
of the table serve as default values for the corresponding \keyw{C*}
and \keyw{TC*} keywords in that same section, but only when axis and
column specifications (including alternate coordinate definitions) use
a time scale listed in Table \ref{table:timescale} or when the
corresponding \keyw{CTYPE} or \keyw{TTYPE} keywords are set to the
value \keyv{TIME}.
Any alternate coordinate specified in a non-recognized time scale
assumes the value of the axis pixels or the column cells, optionally
modified by applicable scaling and/or reference value keywords; see
also Section \ref{timescale}.

\subsection{Precision}
\label{precision}
In order to maintain the precision that is provided by the
HDU, one needs to be careful while processing the information for high
timing precision applications.
Although it is safe to read floating point values in headers and
binary data in double precision, arithmetic performed with those
values may need to be executed with extended precision.
For example, if the header contains:
\begin{verbatim}
  MJDREFI = 1243
  MJDREFF =    0.3746369623
  CRVAL   =    0.0000000111111
  CDELT   =    0.00000000251537257213
\end{verbatim}
then the relative value of the first pixel is:
\begin{verbatim}
  T       = CRVAL + 1*CDELT
          =    0.00000001362647257213
\end{verbatim}
while the final answer, expressed in MJD and performed in quad
precision, is: 
\begin{verbatim}
  TIME    = MJDREFI + MJDREFF + T
          = 1243.374636975926472572130000
\end{verbatim}
The onus is on the application programmer to ensure that applications
maintain their required precision.

\subsection{Labeling}
\label{labeling}

We have observed that there is a confusing variation in the labeling
of time axes in figures and presentations. In particular, the usage of
terms like ``TJD'', ``HJD'', and ``BJD'' is highly ambiguous. Julian and
Modified Julian Date counts do not imply any particular time scale or
any particular reference position. The ``B'' in ``BJD'' raises the
question whether it refers to the reference position \keyv{BARYCENTER} or
the time scale \keyv{TDB}. And an expression like ``BJD$-2400000$''
leaves the reader in doubt whether the value is to be taken literally
or whether the author really meant ``BJD$- 2400000.5$''. 
Authors should be explicit about the times that are posted and we
strongly recommend that they
adopt the following convention for axis labeling:\\
\indent \keyw{JD|MJD(<timescale>;<reference position>)}\\
In order to facilitate the correct labeling we recommend that these
strings be provided in the \keyw{CNAME*} and \keyw{TCNA*}
keywords if possible; for instance:\\
\indent \keyv{TCNAM1  = 'MJD(TDB;Barycenter)'}\\
Also, see the examples of \keyw{TCNA1E} and \keyw{TCNA1F} in Table
\ref{tab:example4}. 

\section{Usage Contexts}
\label{ucontext}
In this section we discuss usage in the contexts to which this
WCS time standard applies; these contexts refer back to
Section~\ref{refterms}.

\subsection{Header Keywords}
The rules governing these keywords are explained in Section
\ref{components} and summarized in Table \ref{table:key}.

\subsection{Time Axis in Images}
\label{images}
Example 1 (Table \ref{tab:example1}) is a data cube in which the $3^{rd}$
axis is time. It is in fact a sequence of 2-D images stacked together.

The rules governing keywords defining the time axis in an image (which
could be a one-dimensional time series or a multi-dimensional
space-time-spectral hypercube) are also dealt with in
Section \ref{components} and summarized in Table \ref{table:key}, but
there are some aspects that require further elaboration as presented
in the following sub-sections.

\subsubsection{Restrictions on Alternate Descriptions}
\label{restrictions}
An image will have at most one time axis as identified by having the
\CTYPE{i} value of \keyv{TIME} or one of the values listed in Table
\ref{table:timescale}. Consequently, as long as the 
axis is identified through \CTYPE{i}, there is no need to have axis
number identification on the global time-related keywords. In addition,
we expressly prohibit the specification of multiple time reference
positions on this axis for alternate time coordinate frames, since
this would give rise to complicated model-dependent non-linear
relations between these frames. Hence, time scales \keyv{TDB} and
\keyv{TCB} (or \keyv{ET}, to its precision) may be specified in the
same image, but cannot be combined 
with any of the first nine time scales in Table \ref{table:timescale};
those first nine can be expressed as linear transformations of
each other, too, provided the reference position remains unchanged.
Time scale \keyv{LOCAL} is by itself, intended for simulations, and
should not be mixed with any of the others.

\subsubsection{\CRVAL{ia}}
The WCS standard requires this keyword to be numeric and cannot be
expressed in ISO-8601 format. Therefore, \CRVAL{ia} is required to
contain the elapsed time in units of \TIMEUNIT{} or \CUNIT{ia}, even
if the zero point of time is specified by \DATEREF{}.

\subsubsection{\CDELT{ia}, \CD{i}{ja} and \PC{i}{ja}}
\label{cdelt}
If the image does not use a matrix for scaling, rotation and shear
(Paper~I~\cite{paper1}),
\CDELT{ia} provides the numeric value for the time interval.

If the \keyv{PC} form of scaling, rotation and shear
(Paper~I~\cite{paper1}) is used, \CDELT{ia} provides the numeric value
for the time interval, and \PC{i}{j}, where $i = j =$ the index of the
time axis (in the typical case of an image cube with axis 3 being
time, $i=j=3$) would take the exact value 1, the default (Paper~I~\cite{paper1}).

When the \CD{i}{j} form of mapping is used, \CD{i}{j} provides the numeric
value for the time interval.

If one of the axes is time and the matrix form is used, then the
treatment of the \PC{i}{ja} (or \CD{i}{ja}) matrices involves at least
a Minkowsky metric and Lorentz transformations (as contrasted with
Euclidean and Galilean).
See Soffel et al.\ (\cite{soffel}) for a full review of the IAU
resolutions concerning space-time coordinate transformations.

Sections \ref{slit} and \ref{untract} describe examples of the use of
these keywords.

\subsubsection{Example of an Image Constructed by a Moving Slit}
\label{slit}

As an example we present a header in Table \ref{tab:example2}
(Example 2) based on a simplified version of a SOHO Coronal Diagnostic
Spectrometer observation from October 1998 (Harris, et
al.\ \cite{harris}).  An image of the Sun is focused onto the entrance slit 
of a stigmatic spectrograph, forming a spectral image on the intensified CCD
detector with wavelength in one direction, and the latitudinal spatial
dimension in the other direction.  A spectrally resolved map of the Sun is
formed by moving the slit from right to left during the observation; thus
different parts (columns) of the data cube are observed at different times.
The example header defines the relations between the different coordinate
systems by specifying a degenerate Time axis that is related to the first
spatial pixel axis through the \PC{4}{2} matrix element.

An alternative approach for the example in Table \ref{tab:example2}
would be to define the time axis as \CTYPE{2}{A}, tying it directly
to the longitude pixel coordinate. However, it is possible to devise a
scenario where this simple alternative approach is not sufficient.  In
the actual observation that Table \ref{tab:example2} is based on, the
slit was tilted relative to solar north, so that the resulting
\PC{i}{j} matrix would have non-trivial values for axes 2 and 3.  If
the data were then rotated to be aligned to solar north, time would be
dependent on both spatial axes, which would be reflected as non-zero
values for both \PC{4}{2} and \PC{4}{3}.

This approach is shown in Example 3 (Table \ref{tab:example3}). One
could still use the alternative description of Time as an alternate
axis on longitude, but in that case it would need its own \PC{2}{jA}
matrix.

\subsubsection{Less Tractable Space-Time Interactions}
\label{untract}

The following example does not have a fool-proof solution, but it may
be instructive. It is derived from the APF telescope at Lick Observatory.
This is a telescope on an azimuth-elevation mount where the guider has
no rotator, so the Celestial WCS changes as the telescope tracks.
The guide camera software can produce movies as 3-D FITS files.

There is no provision for a Celestial WCS which changes as a function
of time (or position), so it is not possible for a FITS file to store
a complete description of the WCS for every frame in a movie
within the context of a single HDU.

However, it is possible to store a WCS for the beginning and end of a movie.
That allows standard FITS WCS viewing programs to give some idea of
the amount of field rotation that happens during a movie.

So the intent of the FITS header in Example 4 (Table
\ref{tab:example6}) is to communicate that alternate WCS \keyv{S} is
valid at the beginning of the exposure and alternate WCS \keyv{R} is
valid at the end of the exposure.

Of course, an alternate approach would be to provide the WCS
information for each frame in a binary table as a separate HDU.
Each row in the table would represent a separate time step and the
columns would contain the corresponding time-dependent WCS parameters
using the Green Bank convention (Pence~\cite{pence2}).
This solution has the benefit of
providing exact WCS information. However, it does require introducing a
separate HDU, whereas the merit of the example in Table \ref{tab:example6}
is that it provides the extremes within the image HDU itself.
In conclusion, these two approaches may be considered complementary
and are not mutually exclusive.

\subsection{Time Columns in Tables}
\label{timecol}
Example 5 (Table \ref{tab:example4}) is part
of the header of an 
event list (a binary table in pixel list mode) with two time
columns. Column 1 carries time in TT, with alternate time coordinate
frames in UTC, TCG, Mission Elapsed Time, Observation Elapsed Time,
MJD, and JD.
Column 20 contains the time stamps in TDB with alternate frames in
TCB and Julian epoch; columns 21 and 22 provide the events' positions. 

The rules governing keywords defining the time in table columns
(pixel as well as vector columns) are
largely dealt with in Section \ref{components} and summarized
in Table \ref{table:key}, but, again, there are some aspects that
require further elaboration.

All times (other than ISO-8601), expressed in a recognized time scale
(see Table \ref{table:timescale}), are relative (to \MJDREF{},
\JDREF{} or \DATEREF{}).
That means that they are
elapsed times and that users have to take care of leap seconds when
using UTC; the unit '\keyv{d}' is defined as 86400 elapsed seconds.
But beware of the following: the reference time values are to be taken
in the time scale specified for the coordinate one is dealing with.
That is why the \keyw{TCRV1A} in the Table \ref{tab:example4} needs
to account for the difference between \keyw{MJDREF}(TT) and
\keyw{MJDREF}(UTC).

Times that are expressed in any other time scale ({\em e.g.}, Mission
Elapsed Time, a common scale) take the values in the table cells at
face value, though they may be modified by applicable keywords such as
\keyw{TCRP*}, \keyw{TCRV*}, and \keyw{TCD*}.

In the context of tables the most important point to keep in
mind is that \TCTYP{n} and/or \TCTY{na} contain the time scale.
However, it should also be pointed out that a binary table column with
\TTYPE{n} = '\keyv{TIME}' and either lacking any \keyi{TC*}{n}
keywords or with \keyi{TC*}{n} = '\keyv{TIME}' will
be controlled by the global keywords listed in Table \ref{table:key}.
This is a common convention in existing files that will still be
compliant with the present standard.

The keywords \keyw{JEPOCH} and \keyw{BEPOCH}, of course, may also be
turned into table columns. However, one should be mindful that they
are implicitly tied to specific time scales and represent absolute
times. Consequently, they have no association with any of the global
keywords.

\subsubsection{Restrictions}
The same restrictions imposed on the image time axis (see Section
\ref{restrictions}) also apply to individual table columns. However,
since one can have more than one column with time information in the
same table, it is possible to mix different time reference positions
and time scales that are not linearly related to each other --
provided that one does not mix these in the same column.

\subsection{Time in Random Groups}
\label{random}
As noted before, the Random Group structure is deprecated;
we include it here for completeness, but this should
not be construed as a statement in support of its continued use.

There are two ways in which time can enter into random group data
(see Greisen \& Harten~\cite{g+h}): as
one of the subarray axes or through a group parameter. In the former
case the situation is identical to that in images and we refer to
Section \ref{images} for the rules. If time is to be transmitted
through a group parameter, it simply means that the
\PTYPE{i} keyword needs to be set to one of
the time scale codes from Table \ref{table:timescale}, just like the
\CTYPE{i}. All the global time reference frame keywords (see Table
\ref{table:key}) apply, just as they would if \CTYPE{i} were set to the
same time scale value, except that there is no possibility of override
since the \keyi{PUNIT}{i}, \keyi{PSYER}{i}, and \keyi{PRDER}{i} keywords
are not defined in the standard.

\subsection{The Time-related Coordinate Axes}
\label{relatedex}
Summarizing the definition of {\em phase}, {\em timelag}, and {\em
  frequency} in Section \ref{related}, we emphasize three key concepts:
\begin{itemize}
\item
\keyv{FREQUENCY} may be used as the abscissa of any spectrum, except
an electo-magnetic spectrum
\item
{\em Phase} can be used as an alternate description of the {\em time}
coordinate; {\em timelag} and {\em frequency} cannot.
\item
The period of a {\em phase} axis may be provided through the keyword
\CPERI{ia} and its equivalents, but only if that period is a constant;
when that is not the case, the period should either be absent or set
to zero.
\end{itemize}

We provide an simple example of a binary table with one {\em time} and two
{\em phase} columns in Example 6 (see Table \ref{tab:example5}).
We readily admit that in this simple example the phase columns can
also be projected onto the time column as linear alternate coordinate
systems. The purpose of the example is to show the use of the {\em
phase} coordinate, not an encouragement to make headers more
complicated than necessary.

\subsection{Use of Besselian and Julian Epochs}
\label{JBepoch}
Use of Besselian and Julian epochs requires special care, as discussed
in Sections \ref{epochs} and \ref{tunit}. Their use is discouraged,
but the fact remains that data exist with a time axis that is so
labeled. We recommend the following convention for identifying these
epochs (illustrated for binary table column 1, but easily translated
to other use cases):
\begin{verbatim}
  DATEREF = '0000-01-01T00:00:00'
  TCUNI1  = 'Ba'    / for Besselian epochs
  TCUNI1  = 'a'     / or 'yr'; for Julian epochs
  TCNAM1  = 'B epoch' / or 'Besselian epoch'
  TCNAM1  = 'J epoch' / or 'Julian epoch'
\end{verbatim}

\begin{acknowledgements}
The authors want to express their deep gratitude and appreciation for
the dedication and tireless efforts of their colleague and friend
Peter Bunclark in moving the work on this paper forward. We received
his last email on 8 December 2008, just two days before his untimely
death. We miss Pete dearly, not only as a great co-author who kept us
on the straight and narrow, but especially as a very good friend. It
was a privilege to have collaborated with him.
We are also very much indebted to former IAU FITS Working Group chair Bill
Pence, who provided valuable comments and kept exhorting us to finally
finish this paper. AHR gratefully acknowledges the many helpful
discussions he had with Jonathan McDowell and the support by NASA
under contract NAS 8-03060 to the Smithsonian Astrophysical
Observatory for operation of the Chandra X-ray Center.
We thank an anonymous referee for helpful comments that resulted in
improved clarity. And we thank Patrick Wallace, Ken Seidelmann, and
George Kaplan for their comments and suggestions.

\end{acknowledgements}

\clearpage \onecolumn

\begin{deluxetable}{l l l l l l l l l}
\tabletypesize{\normalsize}
\tablecolumns{9}
\tablewidth{0pt}
\tablecaption{Keywords for Specifying Time Coordinates
  \label{table:key}}
\tablehead {
 Keyword Description	& Section &
 Global	&
\multicolumn{2}{c}{Images}&
\multicolumn{2}{c}{Table Pixel Columns}&
\multicolumn{2}{c}{Table Vector Columns}
\\
 & & &
Single&
Multiple&
Primary&
Alternate&
Primary&
Alternate
}

\startdata
 \noalign{\vskip 1ex}% 
\multicolumn{9}{c}{\ref{table:key}.a Informational Keywords}\\
 \noalign{\vskip .8ex}% 
\tableline
 \noalign{\vskip .8ex}% 
Date of HDU creation\tablenotemark{a}&\ref{ktvals}	&\keyw{DATE}\\
Date/time of observation	&\ref{ktvals}	&\keyw{DATE-OBS}&&&
\multicolumn{2}{c}{\keyi{DOBS}{n}}&\multicolumn{2}{c}{\keyi{DOBS}{n}}\\
	&	&\keyw{MJD-OBS}&&&
\multicolumn{2}{c}{\keyi{MJDOB}{n}}&\multicolumn{2}{c}{\keyi{MJDOB}{n}}\\
	&	&\keyw{JEPOCH}\\
	&	&\keyw{BEPOCH}\\
Effective date/time of observation	&\ref{ktvals}	&\keyw{DATE-AVG}&&&
\multicolumn{2}{c}{\keyi{DAVG}{n}}&\multicolumn{2}{c}{\keyi{DAVG}{n}}\\
	&	&\keyw{MJD-AVG}&&&
\multicolumn{2}{c}{\keyi{MJDA}{n}}&\multicolumn{2}{c}{\keyi{MJDA}{n}}\\
Start date/time of observation	&\ref{ktvals}	&\keyw{DATE-BEG}\\
	&	&\keyw{MJD-BEG}\\
	&	&\keyw{TSTART}\\
End date/time of observation	&\ref{ktvals}	&\keyw{DATE-END}\\
	&	&\keyw{MJD-END}\\
	&	&\keyw{TSTOP}\\
Net exposure time	&\ref{durations}	&\keyw{XPOSURE}\\
Wall clock exposure time	&\ref{durations}	&\keyw{TELAPSE}\\
 \noalign{\vskip 1ex}% 
\tableline
 \noalign{\vskip 1ex}% 
\multicolumn{9}{c}{\ref{table:key}.b Global Time Reference Frame Keywords and their Optional Context-Specific Override Keywords}\\
 \noalign{\vskip .8ex}% 
\tableline
 \noalign{\vskip .8ex}% 
Time scale\tablenotemark{c}&\ref{timescale}	&\keyw{TIMESYS}&\CTYPE{i}\tablenotemark{d}&\CTYPE{ia}\tablenotemark{d}&\TCTYP{n}\tablenotemark{d}&\TCTY{na}\tablenotemark{d}&\iCTYP{i}{n}&\iCTY{i}{na}
\\
Zero point in MJD&\ref{trefval}	&\MJDREF{}\tablenotemark{b}\\
Zero point in JD&\ref{trefval}	&\JDREF{}\tablenotemark{b}\\
Zero point in ISO-8601&\ref{trefval}	&\DATEREF{}\\
Reference position  &\ref{trefpos}	&\keyw{TREFPOS} &&&\multicolumn{2}{c}{\TRPOS{n}}&\multicolumn{2}{c}{\TRPOS{n}}\\
Reference direction &\ref{trefdir}	&\keyw{TREFDIR} &&&\multicolumn{2}{c}{\TRDIR{n}}&\multicolumn{2}{c}{\TRDIR{n}}\\
Solar System ephemeris&\ref{plephem}	&\keyw{PLEPHEM}&&\\
Time unit&\ref{tunit}	&\keyw{TIMEUNIT}&\CUNIT{i}&\CUNIT{ia}&\TCUNI{n}&\TCUN{na}&\iCUNI{i}{n}&\iCUN{i}{na}\\
Time offset&\ref{tzero}	&\keyw{TIMEOFFS}\\
Absolute Error&\ref{abserror}	&\keyw{TIMSYER}&\CSYER{i}&\CSYER{ia}&\TCSYE{n}&\TCSY{na}&\iCSYE{i}{n}&\iCSY{i}{na}\\
Relative Error&\ref{relerror}	&\keyw{TIMRDER}&\CRDER{i}&\CRDER{ia}&\TCRDE{n}&\TCRD{na}&\iCRDE{i}{n}&\iCRD{i}{na}\\
Time resolution&\ref{timedel}	&\keyw{TIMEDEL}\\
Time location in pixel&\ref{timepixr}	&\keyw{TIMEPIXR}\\
 \noalign{\vskip 1ex}% 
\tableline
 \noalign{\vskip 1ex}% 
\multicolumn{9}{c}{\ref{table:key}.c Additional Context-Specific
  Keywords for the Time Reference Frame}\\
 \noalign{\vskip .8ex}% 
\tableline
 \noalign{\vskip .8ex}% 
Time axis name&\ref{images}	&&\CNAME{i}&\CNAME{ia}&\TCNAM{n}&\TCNA{na}&\iCNAM{i}{n}&\iCNA{i}{na}\\
Time axis reference pixel&\ref{images}	&&\CRPIX{i}&\CRPIX{ia}&\TCRPX{n}&\TCRP{na}&\iCRPX{i}{n}&\iCRP{i}{na}\\
Time axis reference value&\ref{images}	&&\CRVAL{i}&\CRVAL{ia}&\TCRVL{n}&\TCRV{na}&\iCRVL{i}{n}&\iCRV{i}{na}\\
Time scaling&\ref{cdelt}	&&\CDELT{i}&\CDELT{ia}&\TCDLT{n}&\TCDE{na}&\iCDLT{i}{n}&\iCDE{i}{na}\\
Period for temporal phase\tablenotemark{e}&\ref{related}	&&\CPERI{i}&\CPERI{ia}&\TCPER{n}&\TCPR{na}&\iCPER{i}{n}&\iCPR{i}{na}\\
Zero phase time\tablenotemark{e}&\ref{related}	&&\CZPHS{i}&\CZPHS{ia}&\TCZPH{n}&\TCZP{na}&\iCZPH{i}{n}&\iCZP{i}{na}\\
Transformation matrix&\ref{cdelt}	&&\CD{i}{j}&\CD{i}{ja}&\multicolumn{2}{c}{\TC{n}{ka}}&\multicolumn{2}{c}{\iCD{ij}{na}}\\
Transformation matrix&\ref{cdelt}	&&\PC{i}{j}&\PC{i}{ja}&\multicolumn{2}{c}{\TP{n}{ka}}&\multicolumn{2}{c}{\iPC{ij}{na}}\\

\enddata
\tablenotetext{a}{In UTC if the file is constructed on the Earth's surface}
\tablenotetext{b}{These keywords maybe split into an integer
  (\keyw{MJDREFI} or \keyw{JDREFI}) and fractional (\keyw{MJDREFF} or
  \keyw{JDREFF}) part}
\tablenotetext{c}{Use \keyi{PTYPE}{i} in random groups}
\tablenotetext{d}{These keywords may also assume the values \keyv{PHASE}, \keyv{TIMELAG}, or \keyv{FREQUENCY} to specify the corresponding time-related coordinate axes (see Section \ref{related})}
\tablenotetext{e}{Optional; only for use with coordinate type \keyv{PHASE}}
\end{deluxetable}

\begin{deluxetable}{l}
\tabletypesize{\normalsize}
\tablecolumns{1}
\tablewidth{0pt}
\tablecaption{Example 1: Cube with two spatial \& one time axis
\label{tab:example1}}
\tablehead {
 \colhead{
{\tt
123456789 123456789 123456789 123456789 123456789 123456789 123456789 123456789}}}%
\startdata
{\verb++}\\
{\verb+SIMPLE  =                    T / Fits standard+}\\
{\verb+BITPIX  =                  -32 / Bits per pixel+}\\
{\verb+NAXIS   =                    3 / Number of axes+}\\
{\verb+NAXIS1  =                 2048 / Axis length+}\\
{\verb+NAXIS2  =                 2048 / Axis length+}\\
{\verb+NAXIS3  =                   11 / Axis length+}\\
{\verb+DATE    = '2008-10-28T14:39:06' / Date FITS file was generated+}\\
{\verb+OBJECT  = '2008 TC3'           / Name of the object observed+}\\
{\verb+XPOSURE =               1.0011 / Integration time+}\\
{\verb+MJD-OBS =       54746.02749237 / Obs start+}\\
{\verb+DATE-OBS= '2008-10-07T00:39:35.3342'    / Observing date+}\\
{\verb+TELESCOP= 'VISTA   '           / ESO Telescope Name+}\\
{\verb+INSTRUME= 'VIRCAM  '           / Instrument used.+}\\
{\verb+TIMESYS = 'UTC     '           / From Observatory Time System+}\\
{\verb+TREFPOS = 'TOPOCENT'           / Topocentric+}\\
{\verb+MJDREF  =              54746.0 / Time reference point in MJD+}\\
{\verb+RADESYS = 'ICRS    '           / Not equinoctal+}\\
{\verb+CTYPE2  = 'RA---ZPN'           / Zenithal Polynomial Projection+}\\
{\verb+CRVAL2  =     2.01824372640628 / RA at ref pixel+}\\
{\verb+CUNIT2  = 'deg     '           / Angles are degrees+}\\
{\verb+CRPIX2  =               2956.6 / Pixel coordinate at ref point+}\\
{\verb+CTYPE1  = 'DEC--ZPN'           / Zenithal Polynomial Projection+}\\
{\verb+CRVAL1  =     14.8289418840003 / Dec at ref pixel+}\\
{\verb+CUNIT1  = 'deg     '           / Angles are degrees+}\\
{\verb+CRPIX1  =               -448.2 / Pixel coordinate at ref point+}\\
{\verb+CTYPE3  = 'UTC     '           / linear time (UTC)+}\\
{\verb+CRVAL3  =             2375.341 / Relative time of first frame+}\\
{\verb+CUNIT3  = 's       '           / Time unit+}\\
{\verb+CRPIX3  =                  1.0 / Pixel coordinate at ref point+}\\
{\verb+CDELT3  =               13.3629 / Axis scale at reference point+}\\
{\verb+CTYPE3A = 'TT      '           / alternative linear time (TT)+}\\
{\verb+CRVAL3A =             2440.525 / Relative time of first frame+}\\
{\verb+CUNIT3A = 's       '           / Time unit+}\\
{\verb+CRPIX3A =                  1.0 / Pixel coordinate at ref point+}\\
{\verb+CDELT3A =               13.3629 / Axis scale at reference point+}\\
{\verb+OBSGEO-B=             -24.6157 / [deg] Tel geodetic latitute (+=North)+}\\
{\verb+OBSGEO-L=             -70.3976 / [deg] Tel geodetic longitude (+=East)+}\\
{\verb+OBSGEO-H=            2530.0000 / [m]   Tel height above reference ellipsoid+}\\
{\verb+CRDER3  =               0.0819 / random error in timings from fit+}\\
{\verb+CSYER3  =               0.0100 / absolute time error+}\\
{\verb+PC1_1   =     0.999999971570892 / WCS transform matrix element+}\\
{\verb+PC1_2   =     0.000238449608932 / WCS transform matrix element+}\\
{\verb+PC2_1   =    -0.000621542859395 / WCS transform matrix element+}\\
{\verb+PC2_2   =     0.999999806842218 / WCS transform matrix element+}\\
{\verb+CDELT1  =  -9.48575432499806E-5 / Axis scale at reference point+}\\
{\verb+CDELT2  =   9.48683176211164E-5 / Axis scale at reference point+}\\
{\verb+PV1_1   =                   1. / ZPN linear term+}\\
{\verb+PV1_3   =                  42. / ZPN cubic term+}\\
{\verb+END+}\\
{\verb++}\\
\enddata
\end{deluxetable}

\begin{deluxetable}{l}
\tabletypesize{\normalsize}
\tablecolumns{1}
\tablewidth{0pt}
\tablecaption{Example 2: Header extract of an image where Time is
  coupled with Space, built up from individual exposures from
  a stigmatic slit spectrograph stepped across the solar disk
\label{tab:example2}}
\tablehead {
 \colhead{
{\tt
123456789 123456789 123456789 123456789 123456789 123456789 123456789 123456789}}}%
\startdata
{\verb++}\\
{\verb+SIMPLE  =                    T /Written by IDL:  Fri Sep 25 14:01:44 2009       +}\\
{\verb+BITPIX  =                  -32 /Real*4 (floating point)                         +}\\
{\verb+NAXIS   =                    4 /                                                +}\\
{\verb+NAXIS1  =                   20 / Wavelength                                     +}\\
{\verb+NAXIS2  =                  120 / Detector X                                     +}\\
{\verb+NAXIS3  =                  143 / Detector Y                                     +}\\
{\verb+NAXIS4  =                    1 / Time (degenerate)                              +}\\
{\verb+DATE    = '2009-09-25'         /                                                +}\\
{\verb+BUNIT   = 'erg/cm.2/s/sr/Angstrom' /                                            +}\\
{\verb+DATE-OBS= '1998-10-25T16:59:41.823' /                                           +}\\
{\verb+DATEREF = '1998-10-25T16:59:41.823' /                                           +}\\
{\verb+TIMESYS = 'UTC     '           / We will use UTC                                +}\\
{\verb+CTYPE1  = 'WAVE    '           /                                                +}\\
{\verb+CUNIT1  = 'Angstrom'           /                                                +}\\
{\verb+CRPIX1  =              10.5000 /                                                +}\\
{\verb+CRVAL1  =              629.682 /                                                +}\\
{\verb+CDELT1  =           0.11755400 /                                                +}\\
{\verb+CTYPE2  = 'HPLN-TAN'           /                                                +}\\
{\verb+CUNIT2  = 'arcsec  '           /                                                +}\\
{\verb+CRPIX2  =              60.5000 /                                                +}\\
{\verb+CRVAL2  =              897.370 /                                                +}\\
{\verb+CDELT2  =            2.0320000 /                                                +}\\
{\verb+CTYPE3  = 'HPLT-TAN'           /                                                +}\\
{\verb+CUNIT3  = 'arcsec  '           /                                                +}\\
{\verb+CRPIX3  =              72.0000 /                                                +}\\
{\verb+CRVAL3  =             -508.697 /                                                +}\\
{\verb+CDELT3  =            1.6800000 /                                                +}\\
{\verb+CTYPE4  = 'TIME    '           / Might also have been 'UTC'                     +}\\
{\verb+CUNIT4  = 's       '           /                                                +}\\
{\verb+CRPIX4  =              1.00000 /                                                +}\\
{\verb+CRVAL4  =              3147.84 /                                                +}\\
{\verb+CDELT4  =            6344.8602 /                                                +}\\
{\verb+PC1_1   =              1.00000 /                                                +}\\
{\verb+PC1_2   =              0.00000 /                                                +}\\
{\verb+PC1_3   =              0.00000 /                                                +}\\
{\verb+PC1_4   =              0.00000 /                                                +}\\
{\verb+PC2_1   =              0.00000 /                                                +}\\
{\verb+PC2_2   =              1.00000 /                                                +}\\
{\verb+PC2_3   =              0.00000 /                                                +}\\
{\verb+PC2_4   =              0.00000 /                                                +}\\
{\verb+PC3_1   =              0.00000 /                                                +}\\
{\verb+PC3_2   =              0.00000 /                                                +}\\
{\verb+PC3_3   =              1.00000 /                                                +}\\
{\verb+PC3_4   =              0.00000 /                                                +}\\
{\verb+PC4_1   =              0.00000 /                                                +}\\
{\verb+PC4_2   =          -0.00832947 /                                                +}\\
{\verb+PC4_3   =              0.00000 /                                                +}\\
{\verb+PC4_4   =              1.00000 /                                                +}\\
{\verb+END                                                                             +}\\
\enddata
\end{deluxetable}

\begin{deluxetable}{l}
\tabletypesize{\normalsize}
\tablecolumns{1}
\tablewidth{0pt}
\tablecaption{Example 3: Header extract of an image where Time is coupled with
  Space needing a PC matrix
\label{tab:example3}}
\tablehead {
 \colhead{
{\tt
123456789 123456789 123456789 123456789 123456789 123456789 123456789 123456789}}}%
\startdata
{\verb++}\\
{\verb+SIMPLE  =                    T /Written by IDL:  Fri Sep 25 14:01:44 2009       +}\\
{\verb+BITPIX  =                  -32 /Real*4 (floating point)                         +}\\
{\verb+NAXIS   =                    4 /                                                +}\\
{\verb+NAXIS1  =                   20 / Wavelength                                     +}\\
{\verb+NAXIS2  =                  120 / Longitude                                      +}\\
{\verb+NAXIS3  =                  143 / Latitude                                       +}\\
{\verb+NAXIS4  =                    1 / Time (degenerate)                              +}\\
{\verb+DATE    = '2009-09-25'         /                                                +}\\
{\verb+BUNIT   = 'erg/cm^2/s/sr/Angstrom' /                                            +}\\
{\verb+DATE-OBS= '1998-10-25T16:59:41.823' /                                           +}\\
{\verb+DATEREF = '1998-10-25T16:59:41.823' /                                           +}\\
{\verb+TIMESYS = 'UTC     '           / We will use UTC                                +}\\
{\verb+CTYPE1  = 'WAVE    '           /                                                +}\\
{\verb+CUNIT1  = 'Angstrom'           /                                                +}\\
{\verb+CRPIX1  =              10.5000 /                                                +}\\
{\verb+CRVAL1  =              629.682 /                                                +}\\
{\verb+CDELT1  =           0.11755400 /                                                +}\\
{\verb+CTYPE2  = 'HPLN-TAN'           /                                                +}\\
{\verb+CUNIT2  = 'arcsec  '           /                                                +}\\
{\verb+CRPIX2  =              60.5000 /                                                +}\\
{\verb+CRVAL2  =              897.370 /                                                +}\\
{\verb+CDELT2  =            2.0320000 /                                                +}\\
{\verb+CTYPE3  = 'HPLT-TAN'           /                                                +}\\
{\verb+CUNIT3  = 'arcsec  '           /                                                +}\\
{\verb+CRPIX3  =              72.0000 /                                                +}\\
{\verb+CRVAL3  =             -508.697 /                                                +}\\
{\verb+CDELT3  =            1.6800000 /                                                +}\\
{\verb+CTYPE4  = 'TIME    '           / Might also have been 'UTC'                     +}\\
{\verb+CUNIT4  = 's       '           /                                                +}\\
{\verb+CRPIX4  =              1.00000 /                                                +}\\
{\verb+CRVAL4  =              3147.84 /                                                +}\\
{\verb+CDELT4  =            6344.8602 /                                                +}\\
{\verb+PC1_1   =              1.00000 /                                                +}\\
{\verb+PC1_2   =              0.00000 /                                                +}\\
{\verb+PC1_3   =              0.00000 /                                                +}\\
{\verb+PC1_4   =              0.00000 /                                                +}\\
{\verb+PC2_1   =          -0.00128426 /                                                +}\\
{\verb+PC2_2   =             1.000000 /                                                +}\\
{\verb+PC2_3   =          3.50908E-05 /                                                +}\\
{\verb+PC2_4   =              0.00000 /                                                +}\\
{\verb+PC3_1   =          -0.00964133 /                                                +}\\
{\verb+PC3_2   =         -5.13000E-05 /                                                +}\\
{\verb+PC3_3   =             1.000000 /                                                +}\\
{\verb+PC3_4   =              0.00000 /                                                +}\\
{\verb+PC4_1   =              0.00000 /                                                +}\\
{\verb+PC4_2   =          -0.00822348 /                                                +}\\
{\verb+PC4_3   =           0.00109510 /                                                +}\\
{\verb+PC4_4   =              1.00000 /                                                +}\\
{\verb+END                                                                             +}\\
{\verb++}\\
\enddata
\end{deluxetable}

\begin{deluxetable}{l}
\tabletypesize{\normalsize}
\tablecolumns{1}
\tablewidth{0pt}
\tablecaption{
Example 4: Header extract of an image cube where Space is coupled
with Time through rotation, using different CD matrices for the
beginning and end of the observation
\label{tab:example6}}
\tablehead {
 \colhead{
{\tt
123456789 123456789 123456789 123456789 123456789 123456789 123456789 123456789}}}%
\startdata
{\verb++}\\
{\verb+COMMENT      ----------  Globally valid key words  ----------------             +}\\
{\verb+NAXIS   =                    3 / number of data axes                            +}\\
{\verb+NAXIS1  =                  512 / length of data axis 1                          +}\\
{\verb+NAXIS2  =                  512 / length of data axis 2                          +}\\
{\verb+NAXIS3  =                    7 / length of data axis 3                          +}\\
{\verb+DATEUINI= '2012-04-30T04:44:35.001905' / gettimeofday() basis for begin         +}\\
{\verb+DATE-BEG= '2012-04-30T04:44:32.801905' / estimated begin of initial frame       +}\\
{\verb+XPOSURE =                  14. / [s] total exposure duration                    +}\\
{\verb+GEXPTIME=                   2. / [s] duration of one frame                      +}\\
{\verb+DATEUFIN= '2012-04-30T04:44:47.944826' / gettimeofday() basis for end           +}\\
{\verb+DATE-END= '2012-04-30T04:44:47.744826' / estimated end of final frame           +}\\
{\verb+TIMESYS = 'UTC     '           / time scale                                     +}\\
{\verb+DATEREF = '2012-04-30T04:44:32.801905' / time reference                         +}\\
{\verb+COMMENT      ---------- Celestial WCS at begin of movie ----------              +}\\
{\verb+WCSNAMES= 'sky     '           / APF sky coordinates                            +}\\
{\verb+RADESYSS= 'FK5     '           / celestial coordinate reference system          +}\\
{\verb+EQUINOXS=                2000. / reference frame epoch                          +}\\
{\verb+CTYPE1S = 'DEC--TAN'           / coordinate/projection type for WCS axis i=1    +}\\
{\verb+CTYPE2S = 'RA---TAN'           / coordinate/projection type for WCS axis i=2    +}\\
{\verb+CTYPE3S = 'UTC     '           / coordinate/projection type for WCS axis i=3    +}\\
{\verb+CUNIT1S = 'deg     '           / physical unit for WCS axis i=1                 +}\\
{\verb+CUNIT2S = 'deg     '           / physical unit for WCS axis i=2                 +}\\
{\verb+CUNIT3S = 's       '           / physical unit for WCS axis i=3                 +}\\
{\verb+CRPIX1S =        273.459991455 / FITS axis j=1 pixel location of slit           +}\\
{\verb+CRPIX2S =        257.940002441 / FITS axis j=2 pixel location of slit           +}\\
{\verb+CRPIX3S =                  0.5 / FITS axis j=3 pixel for initial photons        +}\\
{\verb+CRVAL1S =     24.7497222222345 / WCS axis i=1 Dec                               +}\\
{\verb+CRVAL2S =     163.903333333308 / WCS axis i=2 RA                                +}\\
{\verb+CRVAL3S =                   0. / WCS axis i=3 UTC offset from DATEREF           +}\\
{\verb+CD1_1S  = 2.01333824032837E-05 / CTM i_j at begin, note difference from end     +}\\
{\verb+CD1_2S  = -2.16670022704079E-05 / CTM i_j at begin, note difference from end    +}\\
{\verb+CD1_3S  =                   0. / CTM i_j from pixel j to WCS i                  +}\\
{\verb+CD2_1S  = 2.16670022704079E-05 / CTM i_j at begin, note difference from end     +}\\
{\verb+CD2_2S  = 2.01333824032837E-05 / CTM i_j at begin, note difference from end     +}\\
{\verb+CD2_3S  =                   0. / CTM i_j from pixel j to WCS i                  +}\\
{\verb+CD3_1S  =                   0. / CTM i_j from pixel j to WCS i                  +}\\
{\verb+CD3_2S  =                   0. / CTM i_j from pixel j to WCS i                  +}\\
{\verb+CD3_3S  =            2.1632744 / CTM i_j UTC time step of movie frames          +}\\
{\verb+COMMENT      ---------- Celestial WCS at end of movie ------------              +}\\
{\verb+WCSNAMER= 'sky@end '           / APF sky coordinates at end                     +}\\
{\verb+RADESYSR= 'FK5     '           / celestial coordinate reference system          +}\\
{\verb+EQUINOXR=                2000. / reference frame epoch                          +}\\
{\verb+CTYPE1R = 'DEC--TAN'           / coordinate/projection type for WCS axis i=1    +}\\
{\verb+CTYPE2R = 'RA---TAN'           / coordinate/projection type for WCS axis i=2    +}\\
{\verb+CTYPE3R = 'UTC     '           / coordinate/projection type for WCS axis i=3    +}\\
{\verb+CUNIT1R = 'deg     '           / physical unit for WCS axis i=1                 +}\\
{\verb+CUNIT2R = 'deg     '           / physical unit for WCS axis i=2                 +}\\
{\verb+CUNIT3R = 's       '           / physical unit for WCS axis i=3                 +}\\
{\verb+CRPIX1R =        273.459991455 / FITS axis j=1 pixel location of slit           +}\\
{\verb+CRPIX2R =        257.940002441 / FITS axis j=2 pixel location of slit           +}\\
{\verb+CRPIX3R =                  7.5 / FITS axis j=3 pixel for final photons          +}\\
{\verb+CRVAL1R =     24.7497222222345 / WCS axis i=1 Dec                               +}\\
{\verb+CRVAL2R =     163.903333333308 / WCS axis i=2 RA                                +}\\
{\verb+CRVAL3R =            14.942921 / WCS axis i=3 UTC offset from DATEREF           +}\\
{\verb+CD1_1R  = 2.00306908641293E-05 / CTM i_j at end, note difference from begin     +}\\
{\verb+CD1_2R  = -2.17619736671195E-05 / CTM i_j at end, note difference from begin    +}\\
{\verb+CD1_3R  =                   0. / CTM i_j from pixel j to WCS i                  +}\\
{\verb+CD2_1R  = 2.17619736671195E-05 / CTM i_j at end, note difference from begin     +}\\
{\verb+CD2_2R  = 2.00306908641293E-05 / CTM i_j at end, note difference from begin     +}\\
{\verb+CD2_3R  =                   0. / CTM i_j from pixel j to WCS i                  +}\\
{\verb+CD3_1R  =                   0. / CTM i_j from pixel j to WCS i                  +}\\
{\verb+CD3_2R  =                   0. / CTM i_j from pixel j to WCS i                  +}\\
{\verb+CD3_3R  =            2.1632744 / CTM i_j UTC time step of movie frames          +}\\
{\verb+END+}\\
{\verb++}\\
\enddata
\end{deluxetable}

\begin{deluxetable}{l}
\tabletypesize{\normalsize}
\tablecolumns{1}
\tablewidth{0pt}
\tablecaption{Example 5: Header extract of a binary table (event list) with two time columns
\label{tab:example4}}
\tablehead {
 \colhead{
{\tt
123456789 123456789 123456789 123456789 123456789 123456789 123456789 123456789}}}%
\startdata
{\verb+COMMENT      ----------  Globally valid key words  ----------------             +}\\
{\verb+TIMESYS = 'TT      '           / Time system                                    +}\\
{\verb+MJDREF  =  50814.000000000000  / MJD zero point for (native) TT (=  1998-01-01) +}\\
{\verb+MJD-BEG =  53516.157939301     / MJD start time in  (native) TT (on 2005-05-25) +}\\
{\verb+MJD-END =  53516.357939301     / MJD stop  time in  (native) TT                 +}\\
{\verb+MJD-OBS =  53516.257939301     / MJD for observation in (native) TT             +}\\
{\verb+MJD-AVG =  53516.257939301     / MJD at mid-observation in (native) TT          +}\\
{\verb+TSTART  =  233466445.95561     / Start time in MET                              +}\\
{\verb+TSTOP   =  233468097.95561     / Stop time in MET                               +}\\
{\verb+TELAPSE =  1652.0              / Wall clock exposure time                       +}\\
{\verb+XPOSURE =  1648.0              / Net exposure time                              +}\\
{\verb+TIMEPIXR=  0.5000000000000     / default                                        +}\\
{\verb+TIMEDEL =  3.2410400000000     / timedel Lev1 (in seconds)                      +}\\
{\verb+TREFPOS = 'TOPOCENT'           / Time is measured at the telescope              +}\\
{\verb+PLEPHEM = 'DE405   '           / SS ephemeris that is used                      +}\\
{\verb+TIMRDER =  1.0000000000000E-09 / Relative error                                 +}\\
{\verb+TIMSYER =  5.0000000000000E-05 / Absolute error                                 +}\\
{\verb+OBSORBIT= 'orbitf315230701N001_eph1.fits' / Orbit ephemeris file                +}\\
{\verb+RADESYS = 'ICRS    '           / Spatial reference system                       +}\\
{\verb++}\\
{\verb+COMMENT      ----------  First Time Column  -----------------------             +}\\
{\verb+TTYPE1  = 'Time    '           / S/C TT corresponding to mid-exposure           +}\\
{\verb+TFORM1  = '2D      '           / format of field                                +}\\
{\verb+TUNIT1  = 's       '                                                            +}\\
{\verb+TCTYP1  = 'TT      '                                                            +}\\
{\verb+TCNAM1  = 'Terrestrial Time'   / This is TT                                     +}\\
{\verb+TCUNI1  = 's       '                                                            +}\\
{\verb+TCRPX1  = 0.0                  / MJDREF is the true zero point for TIME-TT ...  +}\\
{\verb+TCRVL1  = 0.0                  / ...and relative time is zero there             +}\\
{\verb+TCDLT1  = 1.0                  / 1 s is 1 s                                     +}\\
{\verb+TCRDE1  =  1.0000000000000E-09 / Relative error                                 +}\\
{\verb+TCSYE1  =  5.0000000000000E-05 / Absolute error                                 +}\\
{\verb++}\\
{\verb+TCTY1A  = 'UTC     '           / UTC  ELAPSED seconds since MJDREF              +}\\
{\verb+TCNA1A  = 'Coordinated Universal Time' / This is UTC                            +}\\
{\verb+TCUN1A  = 's       '                                                            +}\\
{\verb+TCRP1A  = 0.0                                                                   +}\\
{\verb+TCRV1A  = 63.184               / TT-TAI=32.184 s, TAI-UTC=31 leap seconds       +}\\
{\verb+TCDE1A  = 1.0                                                                   +}\\
{\verb++}\\
{\verb+TCTY1B  = 'TCG     '           / TCG                                            +}\\
{\verb+TCNA1B  = 'Geocentric Coordinate Time' / This is TCG                            +}\\
{\verb+TCUN1B  = 's       '           / still in seconds                               +}\\
{\verb+TCRP1B  = 0.0                  / MJDREF is the reference point                  +}\\
{\verb+TCRV1B  = 0.46184647           / But TCG is already ahead of TT at MJDREF       +}\\
{\verb+TCDE1B  = 1.0000000006969290   / And it keeps running faster                    +}\\
{\verb++}\\
{\verb+TCTY1C  = 'MET     '           / Mission Elapsed Time                           +}\\
{\verb+TCNA1C  = 'Mission Elapsed Time' / This is MET                                  +}\\
{\verb++}\\
{\verb+TCTY1D  = 'OET     '           / Observation Elapsed Time                       +}\\
{\verb+TCNA1D  = 'Observation Elapsed Time' / This is OET                              +}\\
{\verb+TCRV1D  = 0.0                  / Reference pixel: 0 is at: ...                  +}\\
{\verb+TCRP1D  =  233466445.95561     / ... start time in MET                          +}\\
{\verb+TCTY1E  = 'MJD     '           / For listing MJD                                +}\\
{\verb+TCNA1E  = 'MJD(TT;Topocenter)' / This allows a properly labeled MJD axis        +}\\
{\verb+TCUN1E  = 'd       '           / Now in days                                    +}\\
{\verb+TCRP1E  = 0.0                  / MET 0 is the reference point                   +}\\
{\verb+TCRV1E  = 50814.0              / Not surprising, that is MJDREF                 +}\\
{\verb+TCDE1E  = 1.157407407407e-05   / = 1/86400                                      +}\\
{\verb++}\\
{\verb+TCTY1F  = 'JD     '            / For listing JD                                 +}\\
{\verb+TCNA1F  = 'JD(TT;Topocenter)'  / This allows a properly labeled JD axis         +}\\
{\verb+TCUN1F  = 'd       '           / Now in days                                    +}\\
{\verb+TCRP1F  = 0.0                  / MET 0 is the reference point                   +}\\
{\verb+TCRV1F  = 2450814.5            / Not surprising, that is JDREF                  +}\\
{\verb+TCDE1F  = 1.157407407407e-05   / = 1/86400                                      +}\\
{\verb++}\\
{\verb+COMMENT      ----------  Second Time Column  ----------------------             +}\\
{\verb+TTYPE20 = 'Barytime'           / S/C TDB corresponding to mid-exposure          +}\\
{\verb+TFORM20 = '2D      '           / format of field                                +}\\
{\verb+TUNIT20 = 's       '                                                            +}\\
{\verb+TCTYP20 = 'TDB     '                                                            +}\\
{\verb+TRPOS20 = 'BARYCENT'           / Time is measured at the Barycenter             +}\\
{\verb+TRDIR20 = 'EventRA,EventDEC'   / Reference direction is found in cols 21 and 22 +}\\
{\verb+TCNAM20 = 'Barycentric Dynamical Time' / This is TDB                            +}\\
{\verb+TCUNI20 = 's       '                                                            +}\\
{\verb+TCRPX20 = 0.0                  / MJDREF is the true zero point for Barytime ... +}\\
{\verb+TCRVL20 = 0.0                  / ...and relative time is zero there             +}\\
{\verb+TCDLT20 = 1.0                  / 1 s is 1 s                                     +}\\
{\verb+TCRDE20 =  1.0000000000000E-09 / Relative error                                 +}\\
{\verb+TCSYE20 =  5.0000000000000E-05 / Absolute error                                 +}\\
{\verb++}\\
{\verb+TCTY20C = 'TCB     '           / TCB                                            +}\\
{\verb+TCNA20C = 'Barycentric Coordinate Time' / This is TCB                           +}\\
{\verb+TCUN20C = 's       '           / still in seconds                               +}\\
{\verb+TCRP20C = 0.0                  / MJDREF is the reference point                  +}\\
{\verb+TCRV20C = 10.27517360          / But TCB is already ahead of TDB at MJDREF      +}\\
{\verb+TCDE20C = 1.00000001550520     / And it keeps running faster                    +}\\
{\verb++}\\
{\verb+TCTY20G = 'JEPOCH  '           / JEPOCH                                         +}\\
{\verb+TCNA20G = 'Julian Epoch'       / This is a proper Julian epoch                  +}\\
{\verb+TCUN20G = 'a       '           / now in Julian years                            +}\\
{\verb+TCRP20G = 63115200             / 2000-01-01T12:00:00 in MET seconds             +}\\
{\verb+TCRV20G = 2000.0               / J2000.0                                        +}\\
{\verb+TCDE20G = 3.16880878141E-08    / Convert seconds to Julian years                +}\\
{\verb++}\\
{\verb+COMMENT      ----------  RA and Dec of each photon event  ---------             +}\\
{\verb+TTYPE21 = 'EventRA '           / RA of photon event                             +}\\
{\verb+TFORM21 = 'D       '           / format of field                                +}\\
{\verb+TUNIT21 = 'deg     '                                                            +}\\
{\verb+TTYPE22 = 'EventDEC'           / Dec of photon event                            +}\\
{\verb+TFORM22 = 'D       '           / format of field                                +}\\
{\verb+TUNIT22 = 'deg     '                                                            +}\\
{\verb++}\\
{\verb+END+}\\
{\verb++}\\
\enddata
\end{deluxetable}

\begin{deluxetable}{l}
\tabletypesize{\normalsize}
\tablecolumns{1}
\tablewidth{0pt}
\tablecaption{Example 6: Header extract of a binary table with two phase columns
\label{tab:example5}}
\tablehead {
 \colhead{
{\tt
123456789 123456789 123456789 123456789 123456789 123456789 123456789 123456789}}}%
\startdata
{\verb++}\\
{\verb+COMMENT      ----------  Globally valid key words  ----------------             +}\\
{\verb+TIMESYS = 'TT      '           / Time system                                    +}\\
{\verb+TIMEUNIT= 's       '           / Default time unit; applies to TCZPHi, TCPERi   +}\\
{\verb+MJDREFI =  53516               / MJD zero point for TT (integer part)           +}\\
{\verb+MJDREFF =      0.157939301     / MJD zero point for TT (fractional part)        +}\\
{\verb+MJD-BEG =  53516.157939301     / MJD start time in  (native) TT                 +}\\
{\verb+MJD-END =  53516.357939301     / MJD stop  time in  (native) TT                 +}\\
{\verb+MJD-OBS =  53516.257939301     / MJD for observation in (native) TT             +}\\
{\verb+MJD-AVG =  53516.257939301     / MJD at mid-observation in (native) TT          +}\\
{\verb+TSTART  =  0.0                 / Start time in MET                              +}\\
{\verb+TSTOP   =  1652.0              / Stop time in MET                               +}\\
{\verb+TELAPSE =  1652.0              / Wall clock exposure time                       +}\\
{\verb+TREFPOS = 'TOPOCENT'           / Time is measured at the telescope              +}\\
{\verb+PLEPHEM = 'DE405   '           / SS ephemeris that is used                      +}\\
{\verb+TIMRDER =  1.0000000000000E-09 / Relative error                                 +}\\
{\verb+TIMSYER =  5.0000000000000E-05 / Absolute error                                 +}\\
{\verb+OBSORBIT= 'orbitf315230701N001_eph1.fits' / Orbit ephemeris file                +}\\
{\verb++}\\
{\verb+COMMENT      ---------- Time Column  ------------------------------             +}\\
{\verb+TTYPE1  = 'Time    '           / S/C TT                                         +}\\
{\verb+TFORM1  = 'D       '           / format of field                                +}\\
{\verb+TUNIT1  = 's       '                                                            +}\\
{\verb+TCTYP1  = 'TT      '                                                            +}\\
{\verb+TCNAM1  = 'Terrestrial Time'   / This is TT                                     +}\\
{\verb+TCUNI1  = 's       '                                                            +}\\
{\verb+TCRPX1  = 0.0                  / MJDREF is the true zero point for TIME-TT ...  +}\\
{\verb+TCRVL1  = 0.0                  / ...and relative time is zero there             +}\\
{\verb+TCDLT1  = 1.0                  / 1 s is 1 s                                     +}\\
{\verb++}\\
{\verb+COMMENT      ----------  First Phase Column  ----------------------             +}\\
{\verb+TTYPE2  = 'Phase_1 '           / Phase of feature 1                             +}\\
{\verb+TFORM2  = 'D       '           / format of field                                +}\\
{\verb+TUNIT2  = 'turn    '                                                            +}\\
{\verb+TCTYP2  = 'PHASE   '                                                            +}\\
{\verb+TCNAM2  = 'Phase of Feature 1' / Just a name                                    +}\\
{\verb+TCZPH2  = 0.0                  / Phase=0 occurs at MJDREF[IF]                   +}\\
{\verb+TCPER2  = 1652.0               / The period for this phase column               +}\\
{\verb++}\\
{\verb+COMMENT      ----------  Second Phase Column  ---------------------             +}\\
{\verb+TTYPE3  = 'Phase_2 '           / Phase of feature 2                             +}\\
{\verb+TFORM3  = 'D       '           / format of field                                +}\\
{\verb+TUNIT3  = 'turn    '                                                            +}\\
{\verb+TCTYP3  = 'PHASE   '                                                            +}\\
{\verb+TCNAM3  = 'Phase of Feature 2' / Just a name                                    +}\\
{\verb+TCZPH3  = 826.0                / Phase=0 occurs at this offset from MJDREF[IF]  +}\\
{\verb+TCPER3  = 3304.0               / The period for this phase column               +}\\
{\verb++}\\
{\verb+COMMENT      ----------  Observable  ------------------------------             +}\\
{\verb+TTYPE4  = 'Observable'         / Some random quantity                           +}\\
{\verb+TFORM4  = 'D       '           / format of field                                +}\\
{\verb+END+}\\
{\verb++}\\
{\verb++}\\
{\verb++}\\
{\verb+=======================  Data =====================================             +}\\
{\verb++}\\
{\verb+Row     Time       Phase_1   Phase_2   Observable+}\\
{\verb+-------------------------------------------------+}\\
{\verb+  1       0.0        0.0      0.75        10.0 +}\\
{\verb+  2     165.2        0.1      0.80        20.0 +}\\
{\verb+  3     330.4        0.2      0.85        40.0 +}\\
{\verb+  4     495.6        0.3      0.90        80.0 +}\\
{\verb+  5     660.8        0.4      0.95        70.0 +}\\
{\verb+  6     826.0        0.5      0.00        60.0 +}\\
{\verb+  7     991.2        0.6      0.05        50.0 +}\\
{\verb+  8    1156.4        0.7      0.10        40.0 +}\\
{\verb+  9    1486.8        0.9      0.20        20.0 +}\\
{\verb+ 10    1652.0        0.0      0.25        10.0 +}\\
{\verb++}\\
{\verb+End of data+}\\
\enddata
\end{deluxetable}

\appendix
\section{Time Scales}
\label{Appendix}
If one is dealing with high-precision timing, there are more subtle
issues associated with the various time scales that should be considered.
This Appendix provides the necessary information that supplements
Section \ref{timescale} and Table \ref{table:timescale}.
It also provides some background information
on how some of the time scales are realized and how they relate to
each other.
To put this in context, measurement of time is based on the SI
second which has been defined since 1967 as the duration of
9192631770 periods of the radiation corresponding to the transition
between the two hyperfine levels of the ground state of the caesium
133 atom (BIPM~\cite{bipm}). TT is an ideal time scale based on the
SI second that serves as the independent time variable for all
astronomical ephemerides, whereas TAI, as an observationally
determined time scale, differs from an ideal one due to the
limitations inherent in any observational activity. All currently
maintained time scales, with the exception of UT1, are based on the
SI second in the appropriate reference frame.

\subsection{TT and TDT}

TT is defined by Resolution B1.9 of the 24th General Assembly of the
IAU in 2000 at Manchester (IAU~\cite{IAU2000}\footnote
{http://www.iau.org/static/resolutions/IAU2000\_French.pdf}).
This is a re-definition of TT as originally defined by Recommendation
IV of Resolution A4 of the XXIst General Assembly of the IAU in 1991
at Buenos Aires (IAU~\cite{IAU1991}\footnote
{http://www.iau.org/static/resolutions/IAU1991\_French.pdf}).
By that resolution TT was recognized as a better-defined replacement
for TDT.

The initial definition of TT was explained by
Seidelmann \& Fukushima (\cite{SF92}).
For explanation of the redefinition see
Petit in IERS Technical Note 29\footnote
%{http://www.iers.org/nn\_11216/SharedDocs/Publikationen/EN/IERS/Publications/tn/\\TechnNote29/tn29\_\_019,templateId=raw,property=publicationFile.pdf/tn29\_019.pdf}.
% Note that this URL has been forced to break in sensible places
% Remove the '\\' to restore the correct URL
{http://www.iers.org/nn\_11216/IERS/EN/Publications/TechnicalNotes/tn29.html}

Due to the rotation of the Earth (and motion of other bodies), a point
on the surface changes its depth in the gravitational potential of the
solar system.
As noted in Soffel et al. 2003, the proper time experienced by
chronometers on the surface of Earth differs from TT with a diurnal
variation at the picosecond level.

Because TDT never had a satisfactory definition its meaning is
ambiguous at microsecond precision.  For most uses other than
historical tabulation it is more practical to express such time stamps
as TT.

\subsection{TCG and TCB}

TCG and TCB are defined by Recommendation III of Resolution A4 of the
XXIst General Assembly of the IAU in 1991 at Buenos Aires
(IAU~\cite{IAU1991}\footnote
{http://www.iau.org/static/resolutions/IAU1991\_French.pdf}).
Note 4 suggests that precise use of these time scales requires
specification of both the realized time scale (i.e., TAI) and the
theory used to transform from the realized time scale to the
coordinate time scale.  All of the references given above for TT are
also relevant for TCG and TCB.

Given that TT and TCG differ only by a constant rate, a precise value
of TCG is specified by documenting the realization of TT.
Thus we suggest that TCG(TAI) be shorthand for TCG computed from
TT = TAI $+$ 32.184~s or, alternatively, TCG(TT(TAI)).
Likewise, we suggest that TCG(BIPMnn) be shorthand for
TCG(TT(BIPMnn)).

Specifying a precise value for TCB requires documenting a precise
value of TT and additionally a time ephemeris.  A current example of a
time ephemeris is TE405 given by Irwin \& Fukushima (\cite{IF99}).

It is not immediately clear to us how best to express this in a
concise value for the FITS keyword, for there is no guarantee of a
controlled vocabulary for the time ephemerides: nothing
prevents other authors from producing another time ephemeris based on
DE405.
However we may proceed on the assumption that the differences between
any two time ephemerides will be inconsequentially small.  Consequently,
we suggest that TCB(BIPMnn,TE405) be shorthand for
TCB computed from TT(BIPMnn) and TE405.

\subsection{TDB}

TDB is defined by Resolution B3 of the XXVIth General Assembly
of the IAU in 2006 at Prague (IAU~\cite{IAU2006}\footnote
{http://www.iau.org/static/resolutions/IAU2006\_Resol3.pdf}).
This definition is required for microsecond precision.
Note that the reference frame for TDB is barycentric but
it ticks with seconds approximately scaled to match the SI
seconds of TT on the rotating geoid.
For further details on this matter we refer to Chapter 3 in Urban
\& Seidelmann (\cite{urban}).
%; therefore the seconds
%    of TDB are not SI seconds in its own reference frame.}

\subsection{ET}

ET was defined by Clemence (\cite{GC}),
named by Resolution 6 of the 1950 Conference on the Fundamental
Constants of Astronomy held at CNRS in Paris, and adopted by a
recommendation from IAU Commission 4 during the VIIIth General
Assembly in 1952 at Rome.
The definition of ET is based on the works of Newcomb
(\cite{newc95} and \cite{newc98}) and Brown \& Hedrick (\cite{brown}).
At the IAU General Assemblies in 1961 and 1967 Commission 4
designated three improvements on ET named ET0, ET1, and ET2.

Because ET is nonrelativistic its meaning is ambiguous at millisecond
precision.  For most uses other than historical tabulation it is more
practical to express such time stamps as TT or TDB.
For the purposes of historical tabulation we might want to recommend
the use of `ET(ET0)', `ET(ET1)', and `ET(ET2)'.

Starting in 1955 timestamps derived from radio signal distributions based
on (national) atomic clocks became an option; see Section \ref{ut}.

\subsection{TAI}

TAI is defined by BIPM\footnote
{http://www.bipm.org/en/committees/cc/cctf/ccds-1970.html
(Note the 1980 amendment and 
the change implicit by the IAU 1991 Resolution A4)}.

Thus TAI is intended to be the best possible realization of TT,
which means its aim is to be a geocentric coordinate time scale.
Because of deficiencies in the realization,
TAI is only approximately equal to TT $-$ 32.184 s.

TAI is a special case of the atomic time scales because the only valid
realization is the one in Circular T which is published in arrears by
the BIPM.  As such a FITS keyword value of `TAI' should only be used
for timestamps which have been reduced using a chain of chronometers
traceable through Circular T.  TAI should not be used casually.
For example, there are GPS devices which provide time stamps
that claim to be TAI.

TAI should be avoided prior to 1972 because:
\begin{itemize}
  \item TAI had not been authorized until the 14th CGPM in late 1971\footnote
{http://www.bipm.org/en/CGPM/db/14/1/\\
http://www.bipm.org/en/CGPM/db/14/2/}
  \item TAI had not been available for any contemporary time stamping
   mechanisms prior to 1972-01-01
\end{itemize}
TAI should be used with caution prior to 1977 because of the $10^{-12}$
change in rate, and for precision work TAI should always be corrected
using TT(BIPMnn).

For the recording of terrestrial time stamps between 1955 and 1972
we refer to Section \ref{ut}.

\subsection{GPS time}

GPS time is currently defined by the Interface Specification document
``IS-GPS-200F, Revision F''\footnote{http://www.gps.gov/technical/icwg/\#is-gps-200}.
Note that GPS time is aligned to a specific UTC(USNO) epoch, 19~s
behind TAI, with the fractional part matching UTC(USNO) to within a microsecond.

GPS is a convenient source of accurate time.  However, for precise
timestamps it is necessary for applications to know whether the
receiver has implemented the corrections to the satellite clocks
and ionosphere given by the contents of Subframe 1 as documented in
section 20.3.3.3 of IS-GPS-200.

GPS system time should not be used before its date of inception (1980-01-06).

\subsection{UTC}

UTC is defined by ITU-R TF.460 (CCIR \cite{ccir} and ITU \cite{itu}).
It is a time scale that is offset from TAI by an integer number of
SI seconds and required to be within 0.9~s of UT1, Earth rotation
time; this requirement is satisfied by the insertion of leap seconds
as needed, (see also Section \ref{timescale}).
According to the specification the broadcasters are required to match
only to within a millisecond.
Because of the international recommendations and treaty obligations
regarding its use, most national metrology agencies have adopted UTC
and disseminate it as part of their statutory obligation.

UTC should be used with caution prior to 1974 because the meaning of
the name was unknown outside the metrology community.\\
UTC should be used with extreme caution prior to 1972-01-01 because
different contemporary sources of timestamps were providing different
time scales.\\
UTC with its current definition was not available prior to 1972.
Aside from historical tabulations, most terrestrial time stamps prior
to 1972 should be expressed as UT (see Section \ref{ut}) and we
recommend specifically that GMT be interpreted as UT for such dates.\\
UTC should not be used prior to 1960-01-01 because coordination of
broadcast time did not begin until then, and prior to 1961 only
time sources in the US and UK were providing it.

UTC from any source is practical, but in order to provide precision timestamps
one needs to know which realization was used.

UTC from a GPS receiver is also practical, but any tools that are to
provide precision timestamps need
to know whether the receiver has implemented the corrections
given by the contents of Subframes 4 and 5 as documented in section
20.3.3.5 of IS-GPS-200.

\subsection{GMT}

Greenwich Mean Time (GMT) is an ill-defined timescale that nevertheless
continues to persist in popular parlance as well as
scientific papers. Its use is to be discouraged, but if encountered it
should be interpreted as UTC, with the caveat that it is rather
loosely defined as such and any assertions as to the precision of the
time stamps should be regarded with caution.

\subsection{UT}
\label{ut}
The underlying concept for UT originated at the International Meridian
Conference held in Washington in 1884 which defined the Universal Day
as a mean solar day to be reckoned from Greenwich midnight.
UT was initially defined by Newcomb's ``fictitious mean sun''
(Newcomb~\cite{newc95} and \cite{newc98}).
The name Universal Time was established as the subdivision of the
Universal Day by Commission 4 of the IAU at the IIIth General Assembly
in 1928 at Leiden (IAU~\cite{IAU1928}\footnote
{http://www.iau.org/static/resolutions/IAU1928\_French.pdf}).

Most terrestrial time stamps prior to 1972 should be expressed as UT.
For events with time stamps established by radio transmissions
we note that it is possible to use Bulletin Horaire of the BIH to
obtain sub-second precision on one of the time scales here.

Particularly, if high time resolution is required for time stamps
based on radio distributions of atomic clocks between 1955 and 1972
{\em should} be specific about their time distribution source.\\
  Example notations of time stamps before 1972 which were based on
  broadcasts that have not been corrected for propagation time from
  the transmitter:\\
\indent  UT(WWV)\\
\indent  UT(CHU)\\
  Example notations for time stamps before 1972 which were based on
  broadcasts that have been corrected for propagation time from the
  transmitter:\\
\indent  UT(NBS)\\
\indent  UT(NRC)\\
  Example notation for time stamps before 1972 which were based on
  broadcasts that have been corrected to {\em Heure D\'efinitive}
  using {\em Bulletin Horaire}:\\
\indent  UT(BIH)\\
  Example notations for time stamps since 1972-01-01 which were based
  on broadcasts that have not been corrected for propagation time from
  the transmitter:\\
\indent  UTC(JJY)\\
\indent  UTC(DCF77)\\
\indent  UTC(MSF)\\
  Example notations for time stamps since 1972-01-01 which were based
  on broadcasts that have been corrected for propagation time from the
  transmitter:\\
\indent  UTC(NICT)\\
\indent  UTC(PTB)\\
\indent  UTC(NPL)\\
  Example notation for time stamps since 1972-01-01 which have been
  corrected using tabulations published in BIPM {\em Circular T}:\\
\indent  UTC(BIPM)\\
We also encourage the use of a similar convention to denote the flavor
of UT used during that period:\\
\indent UT(UT0)    \\
\indent UT(UT1)    \\
\indent UT(UT2)    \\
    We encourage data publishers to transform their time stamps to a
    uniform time scale such as TT, but we recognize that some data may
    have been acquired by systems for which time offsets were never
    characterized.  Even when those offsets are known the
    transformation may require disproportionate research and
    consultation with the tabulated corrections for radio broadcasts
    published by a national time bureau, and the BIH {\em Bulletin
    Horaire} or BIPM {\em Circular T}.  Therefore this standard
    requires minimally that the original time stamps be transformed
    from a local time zone to UT.

This paper cannot prescribe a historically complete set of
notations for time stamps from radio transmissions and national
laboratories.  When any of these notations for precision time are
used we recommend inclusion of comments describing how the time
stamps should be interpreted.

In exceptional cases of events with time stamps established by
chronometers at observatories with meridian instruments, calibration
is possible to sub-second precision as far back as 1830
(Jordi, et al.\ \cite{jordi}).
\end{document}